\documentclass[journal]{IEEEtran}
\IEEEoverridecommandlockouts
\usepackage{cite}
\usepackage{amsmath,amssymb,amsfonts}
\usepackage{bm}
\usepackage{algorithmic,algorithm}
\usepackage{graphicx}
\usepackage{textcomp}
\usepackage{xcolor}
\usepackage{subfigure}
\usepackage{booktabs}
\usepackage{subcaption}
\def\BibTeX{{\rm B\kern-.05em{\sc i\kern-.025em b}\kern-.08em
T\kern-.1667em\lower.7ex\hbox{E}\kern-.125emX}}
\begin{document}

\title{Wireless Multi-User Interactive Virtual Reality in Metaverse with Edge-Device Collaborative Computing
\author{Caolu Xu, Zhiyong Chen, \emph{Senior Member, IEEE}, Meixia Tao, \emph{Fellow, IEEE}, and Wenjun Zhang, \emph{Fellow, IEEE}}

\thanks{The authors are with the Cooperative Medianet Innovation Center and the Department of Electronic Engineering, Shanghai Jiao Tong University, Shanghai 200240, China (e-mail: \{021034910015, zhiyongchen, mxtao, zhangwenjun\}@sjtu.edu.cn). The paper was presented in part at IEEE GLOBECOM 2023\cite{my_GC}.}}
\maketitle

\begin{abstract}
The immersive nature of the metaverse presents significant challenges for wireless multi-user interactive virtual reality (VR), such as ultra-low latency, high throughput and intensive computing, which place substantial demands on the wireless bandwidth and rendering resources of mobile edge computing (MEC). In this paper, we propose a wireless multi-user interactive VR with edge-device collaborative computing framework to overcome the motion-to-photon (MTP) threshold bottleneck. Specifically, we model the serial-parallel task execution in queues within a foreground and background separation architecture. The rendering indices of background tiles within the prediction window are determined, and both the foreground and selected background tiles are loaded into respective processing queues based on the rendering locations. To minimize the age of sensor information and the power consumption of mobile devices, we optimize rendering decisions and MEC resource allocation subject to the MTP constraint. To address this optimization problem, we design a safe reinforcement learning (RL) algorithm, active queue management-constrained updated projection (AQM-CUP). AQM-CUP constructs an environment suitable for queues, incorporating expired tiles actively discarded in processing buffers into its state and reward system. Experimental results demonstrate that the proposed framework significantly enhances user immersion while reducing device power consumption, and the superiority of the proposed AQM-CUP algorithm over conventional methods in terms of the training convergence and performance metrics.   
\end{abstract}

\section{Introduction}
Immersive communications are envisioned as one of the six usage scenarios for the upcoming sixth generation (6G) mobile communication network\cite{immersion_itu}, and it is expected to profoundly impact how people engage in entertainment, work, and social interactions, etc. One significant development within immersive communications is wireless multi-user interactive virtual reality (VR), which provides seamless transitions between the virtual and real worlds\cite{metaverse}. In the context, wireless multi-user interactive VR has garnered considerable attention from academia and industry.

The demand for immersive experiences poses challenges for mobile networks to support wireless multi-user interactive VR. To prevent user vertigo, it is imperative that the motion-to-photon (MTP) latency remains below 20 ms\cite{mtp_latency}. The stringent MTP threshold, together with high frame rate (exceeding 90 Hz), high-definition and strong-interactivity\cite{meta_quest3, huawei_vr_white_paper}, collectively constitute the immersion in the metaverse. To promote the level of immersion, the mobile edge computing (MEC) technology has been introduced. Whereas, the high-throughput and computing-intensive interactive VR service leads to resource constraints on wireless bandwidth and computing power. Furthermore, the power efficiency of mobile VR devices is also a crucial factor. 
Notably, current state-of-the-art VR head mounted displays (HMDs), such as the Apple Vision Pro and Meta Quest 3, support a relatively short battery life (around 2 hours)\cite{apple_vision_pro, meta_quest3}. Since power efficiency limits the device usage duration on a single charge and effects the weight of batteries, it is necessary to utilize MEC for saving power consumption of mobile devices.

Preferentially, the \emph{interactive} characteristics of interactive VR in this paper are given as follows: \emph{\romannumeral1}) stochastic state update of digital twins by interactive control instructions from sensors; \emph{\romannumeral2}) catastrophic pre-cache overhead in comparison with {$360^{\circ}$} video on account of the additional interactive degrees of freedom, i.e., grid location in the virtual world; \emph{\romannumeral3}) real-time rendering with heterogeneous computational demands from interactive commands. 
Hence, differing from pre-cache chunks as {$360^{\circ}$} video in \cite{Reliability_HetNets, mmwave_compress}, one viable method to cope with the fluctuations in wireless channels and the variations in data volume is to predict the content of future frames and load predictive processed frames into the buffer. 
In particular, trajectory perception and prediction algorithms are evolving rapidly \cite{nature_sensors, viewport_prediction, body_Interaction_prediction}, enabling us to implement prediction in wireless interactive VR. 

\subsection{Related Work}
Prior works have explored how MEC contributes to wireless VR. In \cite{3C_sun}, a joint communication, caching, and computing model for wireless VR video is first proposed, based on determining whether to locally cache field of views (FoVs) in 2D or 3D. \cite{Reliability_HetNets} extends the wireless channel scenarios with mmWave and sub-6 GHz based on \cite{3C_sun}. In \cite{mmwave_compress}, the authors considers the compressed group of pictures under dual-connectivity links and introduced the pipeline of edge-server and user-end operations. The aforementioned works primarily focus on the characteristics of $360^{\circ}$ VR videos that can be pre-cached, making them unsuitable for interactive VR. Some studies emphasize the crucial scheduling role played by MEC among social groups of interactive VR. \cite{infocom_placement} and \cite{JSAC_dynamic_place} investigate dynamic service placement on MEC networks with a distributed architecture. In \cite{VRgame_Zhu}, the resources of MEC are utilized to minimize the average inter-player delay.

To prolong the stringent MTP limitation, several prediction models have been developed to optimize the prediction utility, including the optimization of duration \cite{time_prediction}, encoding rate \cite{FoV_Prediction_Cui}, reflection coefficient \cite{JSAC_prediction_rendering_transmission}, etc. Nevertheless, these works focus on optimizing within a single frame or two adjacent frames, and the longer-term impact of inter-frame resource conflicts caused by prediction is overlooked. Some works consider the inter-frame prediction situation. Proactive and real-time contents correspond to multicast and unicast transmission respectively in \cite{Multicast_prediction}. \cite{cross_frame_prediction} highlights prediction errors in spatial and temporal context. \cite{JSAC_sample} and \cite{task_oriented_meta} implement a system co-design within the metaverse. 
In aforementioned works, a tradeoff between the sensor information freshness \cite{PAoI} and the preprocessing cost arises from prediction. 
Refreshing frames with later sensor information contributes to more precise prediction in terms of FoV,
data size and computational amount \cite{cross_frame_prediction}, but it increases the processing load on the system. 
A general model to address the tradeoff induced by prediction and mitigate the degradation of prediction accuracy due to the sensor aging-of-information (AoI) \cite{PAoI} has not been proposed.

\begin{figure}[t]
\centering{\includegraphics[width=0.5\textwidth]{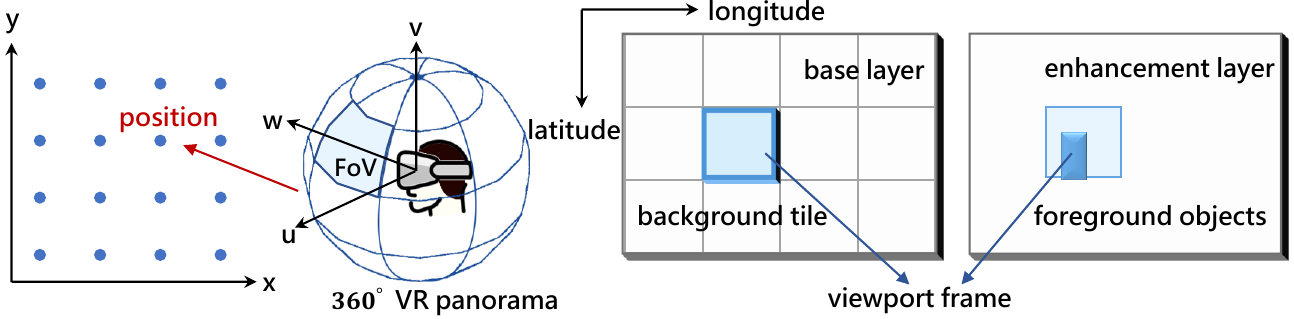}}
\caption{Grid locations in the virtual world. Spatiotemporal VR panorama at location point is divided into FoVs. Merge background tile in the base layer and foreground objects in the enhancement layer to generate viewport frame.}
\label{viewport}
\end{figure}

\begin{figure}[t]
\centering{\includegraphics[width=0.5\textwidth]{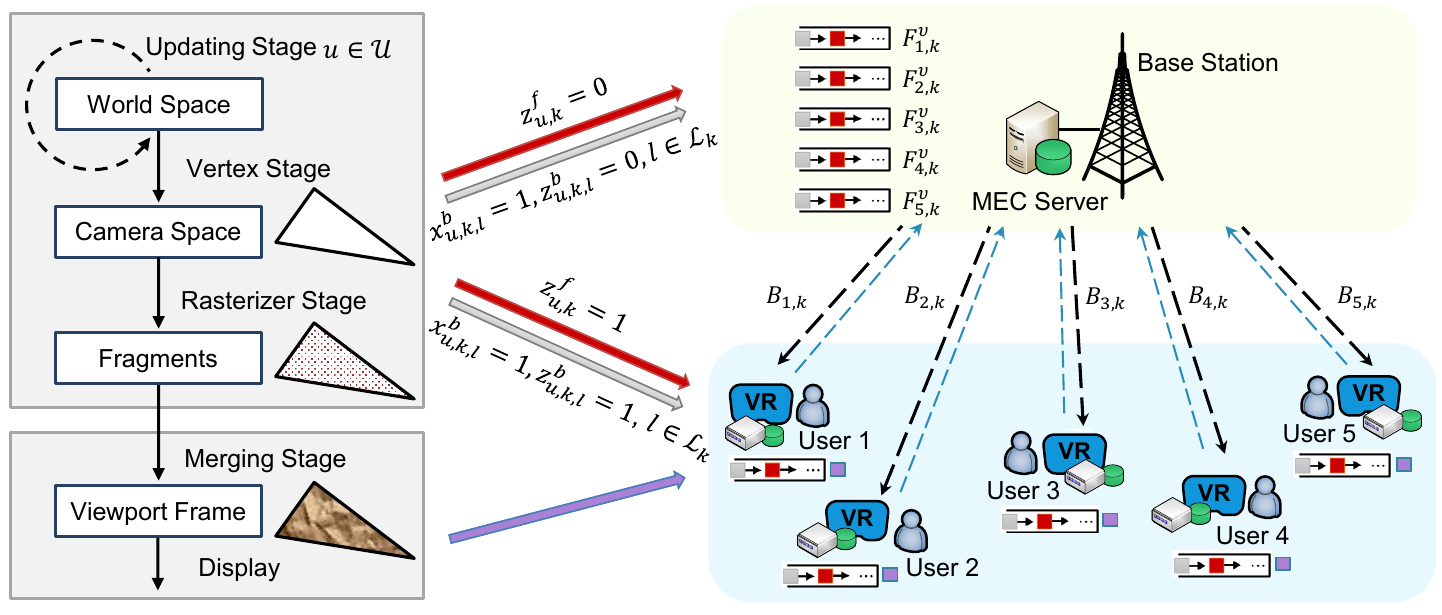}}
\caption{An illustration of the system model at the $k$-th time slot. The left side depicts CG rendering steps.
The right side illustrates wireless multi-user VR with edge-device collaborative computing.}
\label{CG}
\end{figure}

Recently, some studies have started to focus on the features of interactive VR that enhance user experience. A key characteristic pertains to the components of the viewport frame. 
In \cite{back_fore}, a system-level architecture that processes the foreground and background separately is designed. The feasibility of a remote-local rendering structure based on the aforementioned architecture is further verified in \cite{Q-vr}. 
As depicted in Fig. \ref{viewport}, the viewport frame in the virtual world consists of the background environment and foreground objects \cite{my_GC}. The background environment is pre-set, deterministic, not subject to modification. Foreground objects, i.e., digital twins, have uncertain postures and motion states, which are controlled by users. 
The separation of foreground and background layers provides an opportunity for collaborative computing between edge and devices.
To elaborate on how the viewport frame in interactive VR is rendered, an explanation of the computer graphic (CG) rendering process is warranted. Referring to the \emph{OpenGL Graphics System}\cite{segalopengl} in the left side of Fig. \ref{CG}, firstly, foreground objects update vertices into new world space in the updating stage. Then for each user, the vertices of the background and foreground tiles are projected into the first-person camera space in the vertex stage. In the rasterizer stage, the camera space is pixelated according to the display resolution. Color and depth information of the pixels are filled by texture data. Finally, the z-buffer culls primitives which are not facing the camera in the merging stage. According to the aforementioned works, strategically dividing rendering steps and designing the edge-device collaboration workflow are pivotal to break through the MTP threshold bottleneck. 

In the aspect of resource optimization algorithms,  traditional optimization methods encounter difficulties in solving non-convex optimization problems with discrete-continuous optimization variables. The advent of machine learning, particularly deep reinforcement learning (DRL) like deep Q-network (DQN) \cite{nature_dqn} and proximal policy optimization (PPO) \cite{PPO}, has garnered widespread attention. Furthermore, safe reinforcement learning (RL), as a subdomain of RL, has recently emerged as a promising field \cite{safeRL_openai}. Safe RL aims to maximize the expected cumulative reward while adhering to constraints, making it suitable for solving constrained optimization problems \cite{JSAC_prediction_rendering_transmission}. One intuitive approach is using a soft constraint, where the Lagrange multiplier of the constraint is either manually selected or directly learned \cite{JSAC_sample}. Since obtaining the Lagrange multiplier is challenging, this method performs poorly and serves as a baseline. Another approach is to formulate a primal-dual problem that satisfies Slater's condition. Constrained policy optimization (CPO) algorithm \cite{CPO} is designed based on the single-step update in policy gradients. 
Subsequently, the state-of-the-art safe RL method, i.e., constrained updated projection (CUP) \cite{CUP}, extends the theoretical bound of CPO. Due to cross-frame processing tasks, states in RL consist of queue information. It is essential to construct an environment suitable for queues and design a safe RL algorithm to address the rendering decision and resource allocation under the MTP threshold.

\subsection{Contributions}
This paper aims at a prediction, rendering, and communication model for multi-user interactive VR with the insight of edge-device collaborative computing. 
The main contributions are summarized as follows: 
\begin{itemize}
    \item We propose a wireless multi-user interactive VR with edge-device collaborative computing framework to address the ultra-low MTP threshold. This framework comprises three main components: (\romannumeral1) We parallelize several processing tasks to limit the total MTP latency within the threshold, considering serial and parallel timing sequences; (\romannumeral2) We utilize the predictability of background tiles to extend the actual MTP threshold; (\romannumeral3) We pre-cache predicted background tiles to alleviate MTP timeouts resulting from fluctuations in computational load and channel state during the generation and transmission of background tiles across time slots.
    \item 
    We define a quality-of-experience (QoE) metric based on the age of sensor information and the power consumption of mobile devices in multi-user interactive VR. Considering different features of foreground and background tiles in terms of data size, rendering load and predictability, we formulate an optimization problem to maximize the QoE metric while striving to meet the MTP constraint through control decisions and MEC resource allocation. 
    \item We develop an active queue management-constrained updated projection (AQM-CUP) algorithm to solve the optimization problem. First, we reformulate the non-convex problem into a constrained Markov decision process (CMDP). Next, we construct the queue environment and employ AQM to prevent congestion states from propagating. We then derive a suboptimal solution based on the CUP algorithm, which maximizes the QoE metric while adhering to the MTP constraint as much as possible. A simulation environment applied to a real VR dataset for multi-user interactive VR is established. Numerical results demonstrate that the proposed algorithm outperforms traditional methods. 
\end{itemize}

The rest of this paper is organized as follows. Section \ref{System_Model} presents the system model of wireless multi-user interactive VR. Section \ref{problem_formulation} proposes the QoE metric and formulates the optimization problem with MTP threshold. Section \ref{safe_RL_alg} reformulates the problem as CMDP and designs AQM-CUP algorithm to obtain the solution. Simulation results are presented in Section \ref{simulation_results}. Section \ref{conclusion} concludes the paper.

\emph{Notations}: Throughout the paper, caligraphic letters represent sets. $\odot$ is the Hadamard product. $\langle y\rangle^{+}\triangleq\max{(y,0)}$.

\section{System Model}\label{System_Model}
As depicted in Fig. \ref{CG}, we consider a base station (BS) with an MEC server to support several interactive VR users connected via wireless channels. The entire models are already deployed both on the MEC server and on the VR devices. The user set is denoted by $\mathcal{U}=\{1,2,\dots,U\}$.
\subsection{Edge-Device Collaborative Computing Frame}
\begin{figure}[t]
\centering{\includegraphics[width=0.47\textwidth]{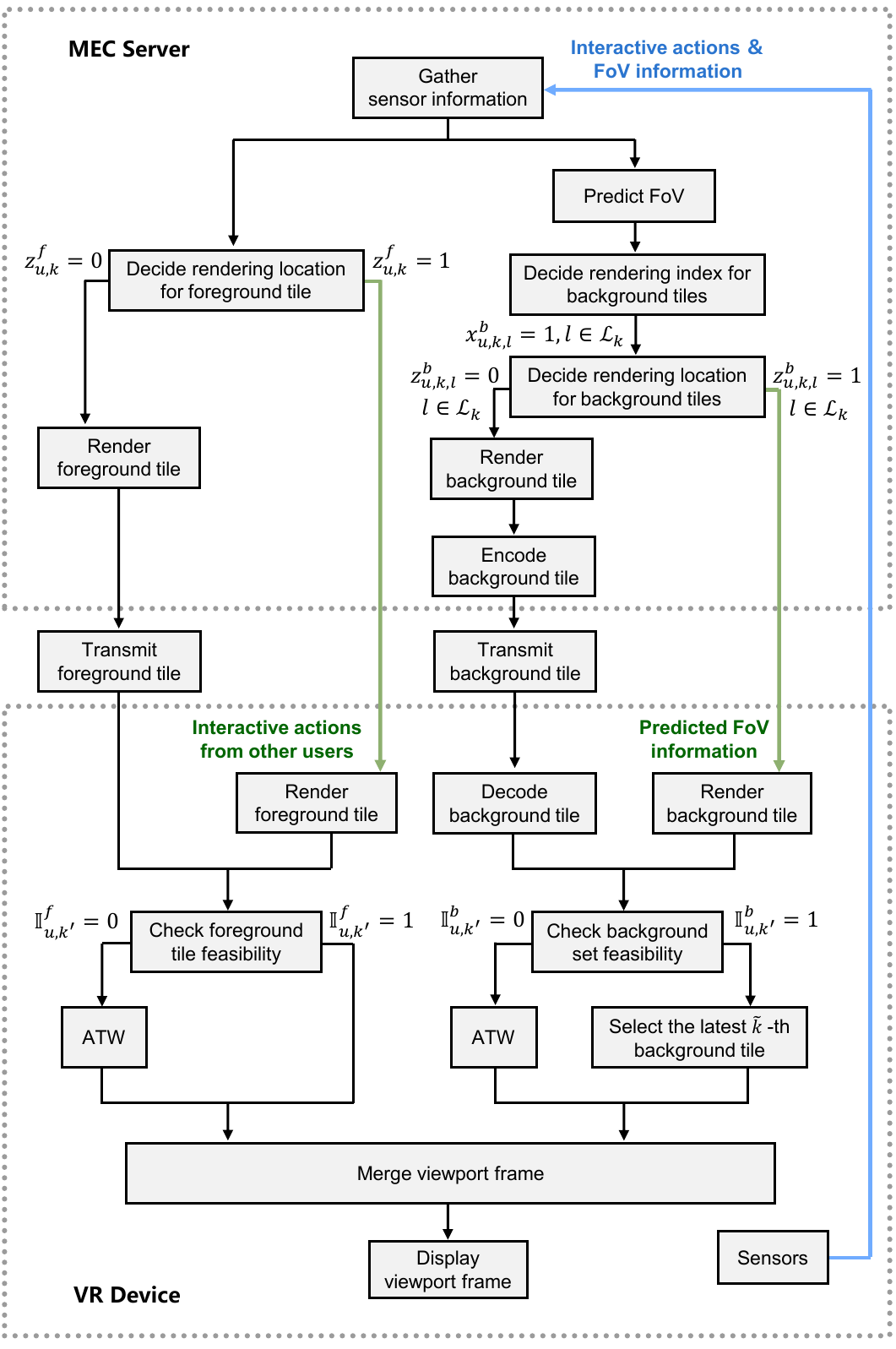}}
\caption{The workflow of interactive VR with edge-device collaborative computing.}
\label{frame}
\end{figure}

Considering a foreground-background separation structure and an edge-device collaborative execution scheme, we design an interactive VR workflow as shown in Fig. \ref{frame}. The requirement of frames per second (FPS) in the interactive VR is denoted by $F$ and the duration of each slot is $\tau=\frac{1}{F}$. During the $k$-th
time slot of user $u$, the workflow consists of the following stages.

  \subsubsection{Upload sensor information}Each user uploads sensor information to MEC. Based on the updated sensor information, the generated data size and computational workload for the foreground and the background are determined.
    
  \subsubsection{Decide rendering index} 
    For the foreground, the obtainable tile at the $k$-th time slot is real-time rendered utilizing the current sensor information.
    For the background, the obtainable window length\footnote{We assume that the background window length remains constant over time, as in \cite{Multicast_prediction}, and only varies with different prediction algorithms.} is denoted by $L\in \mathbb{N}^+$. The FoV, i.e., position and viewing direction, in the index set $\mathcal{L}_{k} \triangleq \{ k, \dots , k+L-1\}$ is attained utilizing the sensor information of the $k$-th time slot. Here, the $k$-th FoV in the $\mathcal{L}_{k}$ corresponds to the current background tile. The FoVs for the $(k+1)$-th and subsequent frames, acquired through prediction algorithms in \cite{viewport_prediction, body_Interaction_prediction}, correspond to future background tiles. 
    If $L=1$, the background cannot be predicted, similar to the foreground. When $L=2$, it signifies that only the background tile of the next frame is predicted, and for $L>2$, it indicates an extended predictive capability. 
Not all background tiles attained at each slot will be rendered since leveraging fresh sensor information to render newly predicted content can enhance the accuracy of the viewport tile, but it simultaneously affects the overhead of this system. Denote $\varphi\in \{f, b\}$ as the tile type (foreground or background). An indicator of whether the $l$-th $\in \mathcal{L}_{k}$ background tile utilizing the sensor information of $k$-th frame for user $u$ will be rendered is denoted as $x^b_{u, k, l}\in \{0,1\}$. If $x^b_{u, k, l} = 1$, this background tile will be added to a rendering queue; otherwise, this tile will not be rendered.

\subsubsection{Determine rendering location} For the foreground, the rendering
location of the foreground tile with the sensor information at the $k$-th time slot for user $u$ is denoted by $z^f_{u,k}\in \{0,1\}$. 
  Similarly, when $x^b_{u,k,l}=1$, $z^b_{u,k,l}\in \{0,1\}$ represents that the background tile of $l$-th frame utilizing the sensor information of $k$-th time slot needs to enter in the MEC server rendering queue $Q^{r_e}_{u}$ or the VR device rendering queue $Q^{r_d}_{u}$. $z^f_{u,k}=0$ (or $z^b_{u,k,l}=0$) indicates the foreground (or the background) is added to  $Q^{r_e}_{u}$. $z^f_{u,k}=1$ (or $z^b_{u,k,l}=1$) indicates the foreground (or the background) is added to $Q^{r_d}_{u}$. 
  Additionally, if $z^f_{u,k}=1$, MEC reports interactive actions\footnote{Compared with a large number of image data, the transmission latency of FoV information and interactive actions in the downlink is negligible.} to the user $u$; if $z^b_{u,k,l}=1$, MEC sends the predicted FoV information of $l$-th frame to user $u$. The real-time foreground tile takes priority over all background tiles and is placed at the front of the rendering queue $Q^{r_e}_{u}$ or $Q^{r_d}_{u}$ when both background and foreground FoVs arrive concurrently.

\subsubsection{Transmit for the tiles rendered by MEC} For the current foreground tile at MEC, the rendered foreground tile is directly added to the wireless transmission queue $Q^{t}_{u}$. The 4K resolution of immersive VR results in a substantial amount of data within a single background tile. The rendered background tiles can be compressed in queue $Q^{c}_{u}$ before entering the wireless transmission queue $Q^{t}_{u}$. Upon reception at the VR device, they are added to the queue $Q^{d}_{u}$ for decompression. $Q^{c}_{u}$, $Q^{d}_{u}$, and $Q^{t}_{u}$ are all first-in-first-out queues.
  
\subsubsection{Check feasibility and generate the viewport frame} New viewport frames need to be generated within the deadline before MTP threshold $T^{\text{MTP}}$. At the current $k$-th time slot of user $u$, the end-to-end latency of the $k^{\prime}$-th viewport frame, where $k^{\prime}\leq k$, is assessed for the MTP threshold violation. For the foreground, $w^f_{u,k^{\prime}}$ denotes the real-time tile utilizing the $k^{\prime}$-th sensor information of user $u$, $\tau^f_{u,k^{\prime}}$ denotes the corresponding end-to-end duration, and $T^f$ denotes the available MTP threshold of the foreground tile with $T^f=T^{\text{MTP}}$. If $\tau^f_{u,k^{\prime}}$ meets the threshold, which is given by
   \begin{equation}\label{tf_th}
  \begin{aligned}
  \tau^f_{u,k^{\prime}} \leq T^f,\ u \in \mathcal{U},
  \end{aligned}
   \end{equation}
$w^f_{u,k^{\prime}}$ is feasible; otherwise, asynchronous time warp (ATW)\cite{ATW} is utilized to generate the $k$-th foreground tile in time. Whether the $k^{\prime}$-th foreground tile of user $u$ comes from the feasible tile $w^f_{u,k^{\prime}}$ or from ATW is denoted by indicator function $\mathbb{I}^f_{u,k^{\prime}}\in\{0,1\}$, where $\mathbb{I}^f_{u,k^{\prime}}=1$ indicates that $\tau^f_{u,k^{\prime}} \leq T^{f}$; otherwise $\mathbb{I}^f_{u,k^{\prime}}=0$. For the background, $w^b_{u,k^{\prime \prime},k^{\prime}}$ denotes the background tile of the $k^{\prime}$-th frame for user $u$ utilizing the $k^{\prime \prime}$-th sensor information, where the $k^{\prime}$-th frame background tile needs to be in the $k^{\prime \prime}$-th obtainable window $k^{\prime}\in \mathcal{L}_{k^{\prime \prime}}$, i.e., $k^{\prime \prime}\in\{\langle k^{\prime}-L\rangle^{+}, \dots, k^{\prime}\}$. Denote $\mathcal{G}_{k^{\prime}}=\{\langle k^{\prime}-L\rangle^{+}, \dots, k^{\prime}\}$. The background feasible set for the $k^{\prime}$-th frame background tile of user $u$ is denoted by $\mathcal W^b_{u,k^{\prime}}$, which is defined as 
  \begin{equation}\label{Sb_uk}
  \begin{aligned}
  \mathcal W^b_{u,k^{\prime}} \triangleq\{w^b_{u,k^{\prime \prime},k^{\prime}}|\ &x^b_{u,k^{\prime \prime},k^{\prime}} = 1, \\&\tau^b_{u,k^{\prime\prime}, k^{\prime}}\leq T^{b}_{k^{\prime\prime}, k^{\prime}}, k^{\prime \prime}\mathrel{\mspace{-3.2mu}}\in\mathrel{\mspace{-3.2mu}}\mathcal{G}_{k^{\prime}}\},
 \end{aligned}
\end{equation}
 where $x^b_{u,k^{\prime \prime},k^{\prime}}$ and $\tau^b_{u,k^{\prime\prime}, k^{\prime}}$ denote the background rendering decision and the end-to-end duration for $w^b_{u,k^{\prime \prime},k^{\prime}}$, respectively. Moreover, $T^{b}_{k^{\prime\prime}, k^{\prime}}$ denotes the available MTP threshold for the $k^{\prime}$-th frame background tile with the $k^{\prime \prime}$-th sensor information, which is given by
\begin{equation}\label{Tb}
 T^{b}_{k^{\prime\prime}, k^{\prime}} = (k^{\prime} - k^{\prime\prime}) \times \tau + T^{\text{MTP}}.
\end{equation}

Hence, \eqref{Sb_uk} indicates that tiles in the $\mathcal W^b_{u,k^{\prime}}$ should be in the rendering schedule and the end-to-end duration must meet the MTP threshold. Before reaching the MTP latency threshold, we check the feasible background set $\mathcal{W}^b_{u,k^{\prime}}$.  
 If $\mathcal{W}^b_{u,k^{\prime}}\neq\varnothing$, we select the feasible background tile of the $k^{\prime}$-th frame with the latest $\Tilde{k}$-th sensor information $w^b_{u,\Tilde{k},k^{\prime}}$. For $\forall u\in \mathcal{U}$, $\Tilde{k}$ is
  \begin{align} \label{kb}
    \Tilde{k}  &= \max\ \{k^{\prime \prime}|\;w^b_{u,k^{\prime \prime},k^{\prime}} \mathrel{\mspace{-2.5mu}}\in\mathrel{\mspace{-2.5mu}}\mathcal{W}^b_{u,k^{\prime}}, k^{\prime \prime}\mathrel{\mspace{-3.5mu}}\in\mathrel{\mspace{-3.5mu}}\mathcal{G}_{k^{\prime}}\}.
 \end{align}
 If $\mathcal{W}^b_{u,k^{\prime}}=\varnothing$, the $k^{\prime}$-th frame background tile of user $u$ has not been rendered in a timely manner and ATW is utilized to generate the background tile instead. $\mathbb{I}^b_{u,k^{\prime}}$ denotes whether the $k^{\prime}$-th background tile of user $u$ comes from the feasible set $\mathcal{W}^b_{u,k^{\prime}}$ or from ATW. If $\mathcal{W}^b_{u,k^{\prime}}\neq\varnothing$, $\mathbb{I}^b_{u,k^{\prime}}=1$; otherwise, $\mathbb{I}^b_{u,k^{\prime}}=0$.
 Finally, the foreground tile is merged with the corresponding background tile to construct the $k^{\prime}$-th viewport frame $w^m_{u,k^{\prime}}$.

\subsection{mmWave Transmission Model}
To address the high-throughput data in interactive VR, we consider mmWave transmission. We consider block fading
and the 3GPP standard for indoor deployment scenarios \cite{3GPP_mmwave} for the mmWave channel. The line of sight (LOS) probability and  non-line of sight (NLOS) probability for the indoor office scenario are denoted by $\rho^{\operatorname{\xi}}_{u,k}, \xi\in \{\text{LOS}, \text{NLOS}\}$, which are distance-dependent functions. Denote $d_{u,k}$ as the distance from the BS to user $u\in \mathcal{U}$ at the $k$-th time slot, then the distance-dependent function $\rho^{\operatorname{LOS}}_{u,k}$ is denoted by $\rho^{\operatorname{LOS}}_{u,k}(d_{u,k})$, and $\rho^{\operatorname{NLOS}}_{u,k}(d_{u,k})=1-\rho^{\operatorname{LOS}}_{u,k}(d_{u,k})$. 
Denote $f^o$ as the carrier frequency of mmWave channel, then the distance and carrier frequency dependent functions of large-scale fading pathloss are denoted by $\ell^{\operatorname{\xi}_{\iota}}_{u,k}(d_{u,k}, f^o), \xi\in \{\text{LOS}, \text{NLOS}\}$ (in dB). The shadowing fading losses for LOS and NLOS links are respectively denoted by $\ell^{\operatorname{\xi}_{\varsigma}}, \xi\in \{\text{LOS}, \text{NLOS}\}$ (in dB). Then, the total pathloss is given by $\ell^{\operatorname{\xi}}_{u,k} = \ell^{\operatorname{\xi}_{\iota}}_{u,k}+\ell^{\operatorname{\xi}_{\varsigma}}$, where $\xi\in \{\text{LOS}, \text{NLOS}\}$. The antenna
gains $g_{u,k}$ between the BS and the user $u$ are considered as a sectorial antenna pattern \cite{mmWave}. $g_{u,k} = (g^\mu)^2$ with the probability of $(\phi/2\pi)^2$, $g_{u,k} = g^\mu g^{\kappa}$ with the probability of $2\phi(2\pi-\phi)/(2\pi)^2$ and $g_{u,k} = (g^{\kappa})^2$ with the probability of $(2\pi-\phi)^2/(2\pi)^2$, where $\phi$ represents the mainlobe beamwidth, $g^\mu$, $g^{\kappa}$ represent the direct gain of main and side lobes, respectively.

The mmWave bandwidth resource allocated by BS to each user at the $k$-th time slot is denoted by $\boldsymbol{B}_k\triangleq(B_{u,k})_{u\in \mathcal{U}}$. The downlink transmission rate $R_{u,k}$ for user $u$ at the $k$-th time slot is given by
\begin{equation}\label{Ruk}
R_{u,k}=B_{u,k} \log _2\left(1+\frac{P\,h_{u,k}\,g_{u,k}}{N_0 B_{u,k} }\right), u \in \mathcal{U},
\end{equation}
where $P$ and $N_0$ respectively represent the transmission power and the noise power spectral density. $h_{u,k}$ is the channel gain and $h_{u,k}=10^{-\ell^{\operatorname{\xi}}_{u,k}/20}, \ \xi\in \{\text{LOS}, \text{NLOS}\}$. 

\subsection{Request and Resource Model}
Denote $\boldsymbol{N}^f_{k}\triangleq(N^f_{u,k})_{u\in\mathcal{U}}$, $\boldsymbol{N}^b_{k}\triangleq(N^b_{u,k^{\dagger},k})_{u\in\mathcal{U}, k^{\dagger}\in \mathcal{G}_k}$ as the rendering loads, where $N^f_{u,k}$ and $N^b_{u,k^{\dagger},k}$ represent the floating-point operations (FLOPs) required to render the real-time foreground tile $w^f_{u,k}$ and the background tile $w^b_{u,k^{\dagger},k}$ within the FoV of user $u$. Referring to \cite{FLOPs_Berkeley},\cite{FLOPs}, the heuristic estimation of FLOPs can be formulated as
\begin{equation}\label{N}
\boldsymbol{N}^{\varphi}_k=\boldsymbol{c}^{v_{\varphi}}_k\odot\boldsymbol{n}^{v_{\varphi}}_k+\boldsymbol{c}^{p_{\varphi}}_k\odot\boldsymbol{n}^{p_{\varphi}}_k,\ \varphi\in\{f,b\},
\end{equation}
where $\boldsymbol{c}^{v_{\varphi}}_k$ and $\boldsymbol{c}^{p_{\varphi}}_k$ represent the complexity of animation and texture, respectively. When $\varphi=f$, $\boldsymbol{c}^{v_f}_k\triangleq (c^{v_{f}}_{u,k})_{u\in\mathcal{U}}$ and $\boldsymbol{c}^{p_f}_k\triangleq (c^{p_{f}}_{u,k})_{u\in\mathcal{U}}$ respectively denote the FLOPs required for one vertex and one pixel of the foreground tile $w^f_{u,k}$. $\boldsymbol{n}^{v_f}_k\triangleq (n^{v_{f}}_{u,k})_{u\in\mathcal{U}}$ and $\boldsymbol{n}^{p_f}_k\triangleq (n^{p_{f}}_{u,k})_{u\in\mathcal{U}}$ denote the number of vertices and pixels in the foreground tile $w^f_{u,k}$. When $\varphi=b$, $\boldsymbol{c}^{v_b}_k\triangleq (c^{v_{b}}_{u,k^{\dagger},k})_{u\in\mathcal{U},k^{\dagger}\in\mathcal{G}_{k}}$ and $\boldsymbol{c}^{p_b}_k\triangleq (c^{p_{b}}_{u,k^{\dagger},k})_{u\in\mathcal{U},k^{\dagger}\in\mathcal{G}_{k}}$ respectively denote the FLOPs required for one vertex and one pixel of the background tile $w^b_{u,k^{\dagger},k}$. $\boldsymbol{n}^{v_b}_k\triangleq (n^{v_{b}}_{u,k^{\dagger},k})_{u\in\mathcal{U},k^{\dagger}\in\mathcal{G}_{k}}$ and $\boldsymbol{n}^{p_b}_k\triangleq(n^{p_{b}}_{u,k^{\dagger},k})_{u\in\mathcal{U},k^{\dagger}\in\mathcal{G}_{k}}$ denote the number of vertices and pixels in the background tile $w^b_{u,k^{\dagger},k}$, respectively. Since $n^{p_{b}}_{u,k^{\dagger},k}$ corresponds to full-screen pixels, it is constant, denoted as $n^{p_{b}}$. 

Denote $\boldsymbol{D}^f_k\triangleq (D^f_{u,k})_{u\in \mathcal{U}}$ where $D^f_{u,k}$ represents the data size of the $k$-th real-time foreground tile $w^f_{u,k}$ for user $u$, which is given by
\begin{equation}\label{Df}
D^f_{u,k} = \varrho\; n^{p_f}_{u,k}, \ u\in\mathcal{U},
\end{equation}
where $\varrho$ represents the number of bits per pixel, $n^{p_f}_{u,k}$ denotes the number of pixels occupied by the foreground tile $w^f_{u,k}$. Especially, $n^{p_f}_{u,k}=0$ indicates that there is no foreground object within the $k$-th FoV of user $u$. 
$w^b_{u,k^{\dagger},k}$ is the background tile of user $u$ for the $k$-th frame utilizing the $k^{\dagger}$-th sensor information. The data size of $w^b_{u,k^{\dagger},k}$ is denoted as $D^b_{u,k^{\dagger},k}$, which is considered to remain constant $D^b=\varrho\,n^{p_{b}}$ across different users and frames \cite{ Q-vr},
\begin{equation}\label{Db}
D^b_{u,k^{\dagger},k}=D^b,\ u\in\mathcal{U}, k^{\dagger}\in\mathcal{G}_{k}.
\end{equation}

The CPU resources at the MEC and users are considered not to be the significant factors causing latency, as stated in \cite{Q-vr}, thus the heterogeneity of CPU resources is not take into account in this paper. For a given background data size $D^b$, the compressing duration at MEC and the decompressing duration at users are assumed to be constant values, denoted as $\Delta^c$ and $\Delta^d$, respectively.
The GPU computation frequency of user $u$ for rendering is denoted by $f^{\nu}_{u}$. The allocated GPU resource at MEC and the downlink transmission rate for each user $u$ during the $k$-th time slot are denoted by $\boldsymbol{F}^{\nu}_k\triangleq(F^{\nu}_{u,k})_{u\in \mathcal{U}}$ and $\boldsymbol{R}_k\triangleq(R_{u,k})_{u\in \mathcal{U}}$. $F^{\nu}_u(t)$ and $R_u(t)$ denote the time-dependent functions of GPU resource and transmission rate for user $u$, respectively.
Hence, $F^{\nu}_u(t)$ and $R_u(t)$ are piecewise constant functions varying over time, which are given by
\begin{align} \label{Fvt}
     F^{\nu}_u(t) = F^{\nu}_{u,k}, \ \text{for } t\in [k\tau, k\tau+\tau), \ k\in \mathbb{N}, \\
     R_u(t) = R_{u,k}, \ \text{for } t\in [k\tau, k\tau+\tau), \ k\in \mathbb{N}.
 \end{align}

\begin{figure*}[t]
\centering{\includegraphics[width=0.87\textwidth]{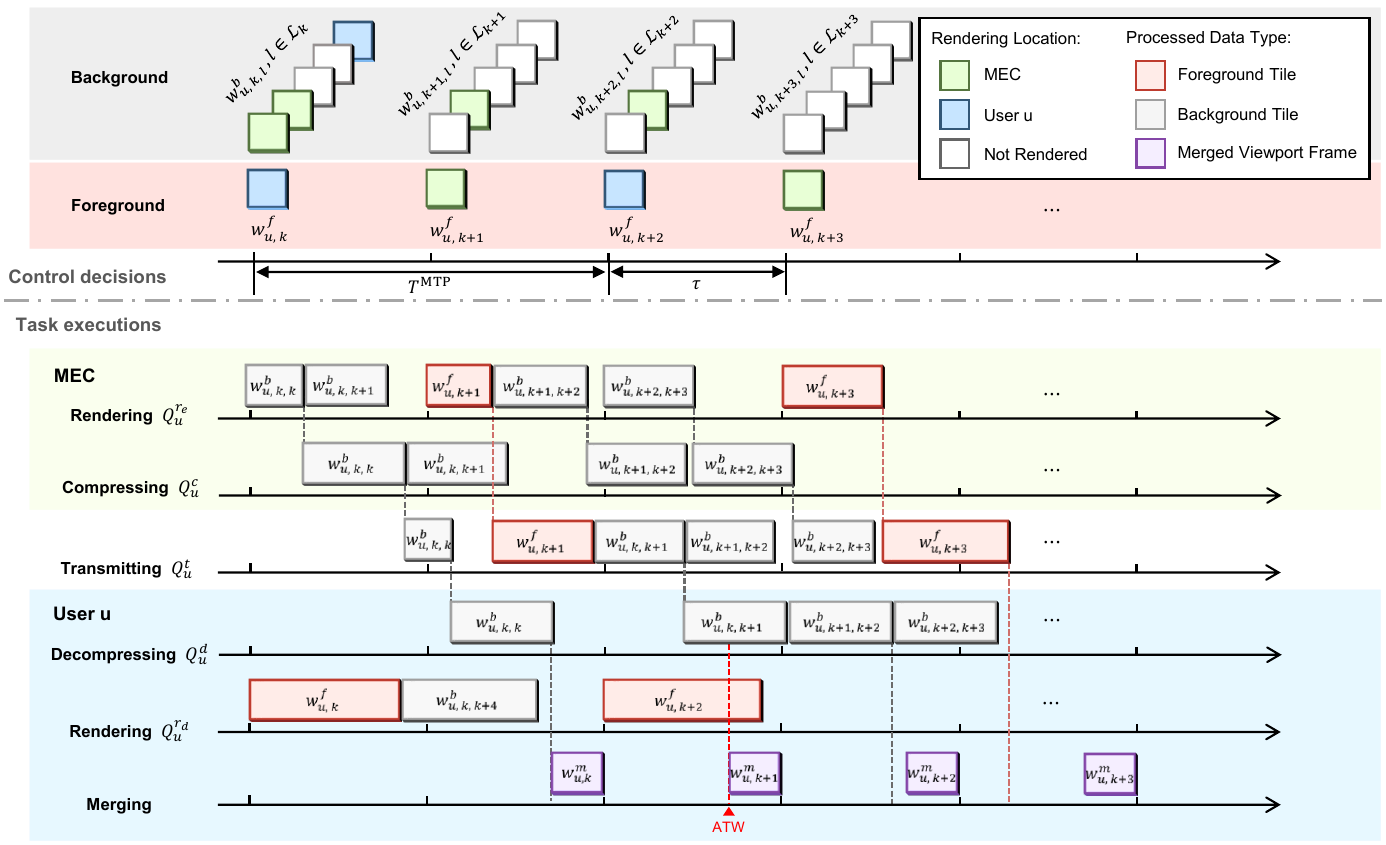}}
	\caption{Timing sequence of control decisions and the corresponding rendering, compressing, transmitting, decompressing and merging tasks in the interactive VR with edge-device collaboration. 
 For simplicity of presentation, (a) all processing queues are empty before the $k$-th time slot, (b) the obtainable window length for the background $L=5$, (c) $\tau$ is set to half of $T^{\text{MTP}}$, i.e., $T^{\text{MTP}} = 20$ ms and $\tau = 10$ ms, corresponding to 100 FPS.}
	\label{mtp latency}
\end{figure*}
 \section{Problem Formulation}\label{problem_formulation}
 \subsection{QoE Metric}
 \subsubsection{Age of sensor information}
 To tradeoff between the freshness of sensor information, i.e. AoI, and the preprocessing cost stemming from prediction, we define a QoE metric $\varkappa_{u,k}$ of the merged $k$-th viewport frame of user $u$ as following
\begin{align} \label{tau_qoe}
     \varkappa_{u,k}=\mathbb{I}_{u,k}(k - k^{\star}) \tau+ (1-\mathbb{I}_{u,k})T,
 \end{align}
where $\mathbb{I}_{u,k}=\mathbb{I}^f_{u,k}\cdot \mathbb{I}^b_{u,k}\in \{0,1\}$ denotes 
whether the $k$-th viewport frame undergoes ATW. $\mathbb{I}_{u,k}=1$ indicates the $k$-th viewport frame of user $u$ is newly rendered without ATW. 
In (\ref{Sb_uk})\text{-}(\ref{kb}), substituting $k^{\prime\prime}=k^{\dagger}$, $k^{\prime}=k$ and $\Tilde{k}=k^{\star}$, $k^{\dagger}$ is the time slot index of the sensor information where the $k$-th frame background tile needs to be in its obtainable window $k\in \mathcal{L}_{k^{\dagger}}$, i.e., the range of $k^{\dagger}$ is denoted as $\mathcal{G}_{k}=\{\langle k-L\rangle^{+}, \dots, k\}$ and $k^{\dagger}\in \mathcal{G}_{k}$,  $k^{\star}$ is the latest index of sensor information among the feasible background tile set $\mathcal{W}^b_{u,k}$.
A larger $k^{\star}$ implies fresher sensor information, contributing to more precise prediction \cite{cross_frame_prediction}, but it results in smaller available MTP threshold $T^b_{k^{\star},k}$, i.e., in (\ref{Tb}), substituting $k^{\prime\prime}=k^{\star}$ and $k^{\prime}=k$, and more stringent requirements for system. 
$\mathbb{I}_{u,k}=0$ indicates the viewport frame is generated by ATW. Then $\varkappa_{u,k}$ in (\ref{tau_qoe}) is a significantly large positive value $T$, and the information freshness is severely compromised.
\subsubsection{Power consumption of devices}
The QoE metric of energy consumption at the interactive VR device for generating the $k$-th viewport of user $u$ is denoted as $\varepsilon_{u,k}$, which consists the energy consumption for decompressing the candidate background tiles of the $k$-th frame, i.e., $\varepsilon^d_{u,k}$, locally rendering the $k$-th foreground and corresponding background tiles, i.e., $\varepsilon^{r_d}_{u,k}$, the merging energy, i.e., $\varepsilon^m_{u,k}$. Since each user merges one frame at each time slot and the energy of merging single frame can be assumed to be constant, $\varepsilon^m_{u,k}$ is fundamental and not subject to optimization. Hence, $\varepsilon^m_{u,k}$ is not involved in metric $\varepsilon_{u,k}$. Consequently, $\varepsilon_{u,k}$ is given by 
\begin{align}\label{energy_qoe}
\varepsilon_{u,k}=\varepsilon^d_{u,k}+\varepsilon^{r_d}_{u,k},
\end{align}
where $\varepsilon^d_{u,k}=\left(\sum\nolimits_{k^{\dagger}\in\mathcal{G}_{k}}x^b_{u,k^{\dagger},k}(1-z^b_{u,k^{\dagger},k})\right)\varepsilon^d$. Constant $\varepsilon^d$ represents the energy consumption of decompressing per background tile $D^b$ per user. Denote $\beta$ as a coefficient related to the VR device hardware, and $\varepsilon^{r_d}_{u,k}=\beta\Big( z^f_{u,k}N^{f}_{u,k}+\sum\nolimits_{k^{\dagger}\in\mathcal{G}_{k}} x^b_{u,k^{\dagger},k}z^b_{u,k^{\dagger},k}N^{b}_{u,k^{\dagger},k}\Big)(f^{\nu}_u)^2$. 
\subsection{MTP Latency Model}
Considering the queuing delays in each process and the serial-parallel relationship of executions, we develop the MTP model based on the timing sequence as depicted in Fig. \ref{mtp latency}. The MTP latency of the $k$-th frame of user $u$ is primarily composed of the following components.

\subsubsection{Rendering foreground tile locally} When $z^f_{u,k}=1$, the foreground tile $w^{f}_{u,k}$ is rendered at the VR device of user $u$. The arrival time that the $k$-th sensor information gets into the rendering queue of user $u$, i.e., $Q^{r_d}_u$, is denoted by $I^{r_{d,f}}_u$, and $I^{r_{d,f}}_u=k\tau$. The latest departure time of the queue $Q^{r_d}_u$ is denoted by $O^{r_d}_u$. The total duration $\tau^{r_{d,f}}_{u,k}$ in $Q^{r_d}_u$ is given by 
\begin{equation}\label{t_rdf}
\tau^{r_{d,f}}_{u,k}=\langle O^{r_d}_u- I^{r_{d,f}}_u\rangle^{+}+\frac{N^f_{u,k}}{f^{\nu}_{u}}.
\end{equation}
where $\langle O^{r_d}_u- I^{r_{d,f}}_u\rangle^{+}$ and $\frac{N^f_{u,k}}{f^{\nu}_{u}}$ are the waiting duration and the service duration for the foreground tile $w^f_{u,k}$ in the queue $Q^{r_d}_u$. 
Then, the latest departure time of the $Q^{r_d}_u$ is updated as $O^{r_d}_u= I^{r_{d,f}}_u+\tau^{r_{d,f}}_{u,k}$.

\subsubsection{Rendering foreground tile remotely}
When $z^f_{u,k}=0$, the foreground tile $w^{f}_{u,k}$ is rendered at MEC. The arrival time of the $k$-th sensor information into the MEC rendering queue $Q^{r_e}_u$ is denoted by $I^{r_{e,f}}_u$, where $I^{r_{e,f}}_u=k\tau$. The latest departure time of the queue $Q^{r_e}_u$ is denoted by $O^{r_e}_u$. The total duration $\tau^{r_{e,f}}_{u,k}$ in $Q^{r_e}_u$ is given by
\begin{equation}\label{t_ref}
\tau^{r_{e,f}}_{u,k}=\langle O^{r_e}_u- I^{r_{e,f}}_u\rangle^{+}+\Delta^{r_{e,f}}_{u,k},
\end{equation}
where $\langle O^{r_e}_u- I^{r_{e,f}}_u\rangle^{+}$ and $\Delta^{r_{e,f}}_{u,k}$ are respectively the waiting duration and the service duration for the foreground tile $w^{f}_{u,k}$ in the queue $Q^{r_e}_u$, and $\Delta^{r_{e,f}}_{u,k}$ is the solution to the following equation
\begin{equation}
\int_{I^{r_{e,f}}_u}^{I^{r_{e,f}}_u+\Delta^{r_{e,f}}_{u,k}} F^{\nu}_u(t) \,dt=N^f_{u,k}.
\end{equation}
Then, the latest departure time of the queue $Q^{r_e}_u$ is updated as $O^{r_e}_u= I^{r_{e,f}}_u+\tau^{r_{e,f}}_{u,k}$.

The foreground tile $w^{f}_{u,k}$ leaving the rendering queue $Q^{r_e}_u$ enters the transmission queue $Q^t_u$. The arrival time that $w^{f}_{u,k}$ gets into $Q^t_u$ is denoted by $I^{t_{f}}_u$, and $I^{t_{f}}_u=O^{r_e}_u$. The latest departure time of the queue $Q^t_u$ is denoted by $O^{t}_u$. The total duration $\tau^{t_{f}}_{u,k}$ in $Q^{t}_u$ is given by
\begin{equation}\label{t_tf}
\tau^{t_{f}}_{u,k}=\langle O^{t}_u- I^{t_{f}}_u\rangle^{+}+\Delta^{t_{f}}_{u,k},
\end{equation}
where $\langle O^{t}_u-I^{t_{f}}_u\rangle^{+}$ and $\Delta^{t_{f}}_{u,k}$ are respectively the waiting duration and the service duration for the foreground tile $w^{f}_{u,k}$ in the queue $Q^{t}_u$, and $\Delta^{t_{f}}_{u,k}$ is the solution to the following equation 
\begin{equation}
\int_{I^{t_{f}}_u}^{I^{t_{f}}_u+\Delta^{t_{f}}_{u,k}} R_u(t) \,dt=D^f_{u,k}.
\end{equation}
Then, the latest departure time of the queue $Q^{t}_u$ is updated as $O^{t}_u= I^{t_{f}}_u+\tau^{t_{f}}_{u,k}$.

When $x^b_{u,k^{\dagger},k}=1$, the background tile $w^b_{u,k^{\dagger},k}$ is rendered; otherwise, $w^b_{u,k^{\dagger},k}$ in the $k^{\dagger}$-th obtainable window will not be rendered. For $x^b_{u,k^{\dagger},k}=1$, the rendering location for $w^b_{u,k^{\dagger},k}$ is further determined by the decision $z^b_{u,k^{\dagger},k}$.

\subsubsection{Rendering background tile locally} When $z^b_{u,k^{\dagger},k}=1$, the background tile $w^{b}_{u,k^{\dagger},k}$ is rendered at the VR device of user $u$. The arrival time that the $k^{\dagger}$-th sensor information enters the rendering queue $Q^{r_d}_u$ is denoted by $I^{r_{d,b}}_u$, and $I^{r_{d,b}}_u=k^{\dagger}\tau$. The total duration $\tau^{r_{d,b}}_{u,k^{\dagger},k}$ in $Q^{r_d}_u$ is 
\begin{equation}\label{t_rdb}
\tau^{r_{d,b}}_{u,k^{\dagger},k}=\langle O^{r_d}_u- I^{r_{d,b}}_u\rangle^{+}+\frac{N^b_{u,k^{\dagger},k}}{f^{\nu}_{u}}.
\end{equation}
where $\langle O^{r_d}_u- I^{r_{d,b}}_u\rangle^{+}$ and $\frac{N^b_{u,k^{\dagger},k}}{f^{\nu}_{u}}$ are the waiting duration and the service duration for the background tile $w^{b}_{u,k^{\dagger},k}$ in the queue $Q^{r_d}_u$. 
Then, the latest departure time of the $Q^{r_d}_u$ is updated as $O^{r_d}_u= I^{r_{d,b}}_u+\tau^{r_{d,b}}_{u,k^{\dagger},k}$.

\subsubsection{Rendering background tile remotely} Likewise, when $z^b_{u,k^{\dagger},k}=0$, the background tile $w^{b}_{u,k^{\dagger},k}$ is rendered at the MEC. The arrival time that the $k^{\dagger}$-th sensor information enters the rendering queue $Q^{r_e}_u$ is denoted by $I^{r_{e,b}}_u$, and $I^{r_{e,b}}_u=k^{\dagger}\tau$. The total duration $\tau^{r_{e,b}}_{u,k^{\dagger},k}$ in $Q^{r_e}_u$ is
\begin{equation}\label{t_reb}
\tau^{r_{e,b}}_{u,k^{\dagger},k}=\langle O^{r_e}_u- I^{r_{e,b}}_u\rangle^{+}+\Delta^{r_{e,b}}_{u,k^{\dagger},k},
\end{equation}
where $\langle O^{r_e}_u- I^{r_{e,b}}_u\rangle^{+}$ and $\Delta^{r_{e,b}}_{u,k^{\dagger},k}$ are respectively the waiting duration and the service duration for the background tile $w^{b}_{u,k^{\dagger},k}$ in the queue $Q^{r_e}_u$, and $\Delta^{r_{e,b}}_{u,k^{\dagger},k}$ is the solution to the following equation
\begin{equation}
\int_{I^{r_{e,b}}_u}^{I^{r_{e,b}}_u+\Delta^{r_{e,b}}_{u,k^{\dagger},k}} F^{\nu}_u(t) \,dt=N^b_{u,k^{\dagger},k}.
\end{equation}
Then, the latest departure time of the queue $Q^{r_e}_u$ is updated as $O^{r_e}_u= I^{r_{e,b}}_u+\tau^{r_{e,b}}_{u,k^{\dagger},k}$.

The background tile $w^{b}_{u,k^{\dagger},k}$ leaving the rendering queue $Q^{r_e}_u$ is compressed in the queue $Q^c_u$. The arrival time that $w^{b}_{u,k^{\dagger},k}$ gets into $Q^c_u$ is denoted by $I^{c}_u$, and $I^{c}_u=O^{r_e}_u$. The latest departure time of the queue $Q^c_u$ is denoted by $O^{c}_u$. The total duration $\tau^{c}_{u,k^{\dagger},k}$ in $Q^c_u$ is
\begin{equation}\label{t_c}
\tau^{c}_{u,k^{\dagger},k}=\langle O^{c}_u- I^{c}_u\rangle^{+}+\Delta^{c},
\end{equation}
where $\langle O^{c}_u- I^{c}_u\rangle^{+}$ and $\Delta^{c}$ are respectively the waiting duration and the service duration for the background tile $w^{b}_{u,k^{\dagger},k}$ in the compressing queue $Q^{c}_u$.
The latest departure time of the queue $Q^{c}_u$ is updated as, $O^{c}_u= I^{c}_u+\tau^{c}_{u,k^{\dagger},k}$. Currently, the background data size is compressed to $\alpha D^b$ and $\alpha$ is the compression ratio.

The compressed background tile $w^{b}_{u,k^{\dagger},k}$
then proceeds to enter the transmission queue $Q^t_u$. The arrival time that $w^{b}_{u,k^{\dagger},k}$ gets into $Q^t_u$ is denoted by $I^{t_{b}}_u$, and $I^{t_{b}}_u=O^{c}_u$. The total duration $\tau^{t_{b}}_{u,k^{\dagger},k}$ in $Q^{t}_u$ is
\begin{equation}\label{t_tb}
\tau^{t_{b}}_{u,k^{\dagger},k}=\langle O^{t}_u- I^{t_{b}}_u\rangle^{+}+\Delta^{t_{b}}_{u,k^{\dagger},k},
\end{equation}
where $\langle O^{t}_u- I^{t_{b}}_u\rangle^{+}$ and $\Delta^{t_{b}}_{u,k^{\dagger},k}$ are respectively the waiting duration and the service duration for the background tile $w^{b}_{u,k^{\dagger},k}$ in the queue $Q^{t}_u$, and $\Delta^{t_{b}}_{u,k^{\dagger},k}$ is the solution to the following equation
\begin{equation}
\int_{I^{t_{b}}_u}^{I^{t_{b}}_u+\Delta^{t_{b}}_{u,k^{\dagger},k}} R_u(t) \,dt=\alpha D^b.
\end{equation}
Then, the latest departure time of the queue $Q^{t}_u$ is updated as $O^{t}_u= I^{t_{b}}_u+\tau^{t_{b}}_{u,k^{\dagger},k}$.

When the VR device receives $w^{b}_{u,k^{\dagger},k}$, the compressed background tile is added to the decompressing queue $Q^{d}_u$. 
The arrival time that $w^{b}_{u,k^{\dagger},k}$ gets into $Q^d_u$ is denoted by $I^{d}_u$, and $I^{d}_u=O^{r_d}_u$. The latest departure time of $Q^d_u$ is denoted by $O^{d}_u$. The total duration $\tau^{d}_{u,k^{\dagger},k}$ in $Q^d_u$ is
\begin{equation}\label{t_d}
\tau^{d}_{u,k^{\dagger},k}=\langle O^{d}_u-I^{d}_u\rangle^{+}+\Delta^{d},
\end{equation}
where $\langle O^{d}_u- I^{d}_u\rangle^{+}$ and $\Delta^{d}$ are respectively the waiting duration and the service duration for the background tile $w^{b}_{u,k^{\dagger},k}$ in the decompressing queue $Q^{d}_u$.
The latest departure time of $Q^{d}_u$ is updated as $O^{d}_u= I^{d}_u+\tau^{d}_{u,k^{\dagger},k}$. 

The merging stage merely relies on depth information and is not the primary source of latency\cite{Q-vr}. Denote the merging duration as $\tau^{m}_{u,k}$, which is assumed to be constant $\tau^{m}$ for all $u\in\mathcal{U}$ and all the time slots.
Accordingly, the end-to-end MTP latency of the foreground tile $w^f_{u,k}$ can be expressed as
\begin{equation}\label{tf}
\tau^f_{u,k}=\left[z^f_{u,k}\ \tau^{r_{d,f}}_{u,k}+\big(1-z^f_{u,k}\big)\big(\tau^{r_{e,f}}_{u,k}+\tau^{t_f}_{u,k}\big)\right]+\tau^m,
\end{equation}
in (\ref{tf_th}) substituting $k^{\prime} = k$, if $\tau^f_{u,k}$ meets the threshold, $w^f_{u,k}$ is feasible and $\mathbb{I}^f_{u,k}=1$. 

For the background tile $w^b_{u,k^{\dagger},k}$ with $x^b_{u,k^{\dagger},k}=1$, the overall end-to-end MTP latency can be formulated as
\begin{align}\label{tb}
\tau^b_{u,k^{\dagger},k}=\Big[z^b_{u,k^{\dagger},k}\ \tau^{r_{d,b}}_{u,k^{\dagger},k}+
\big(1-z^b_{u,k^{\dagger},k}\big)\big(\tau^{r_{e,b}}_{u,k^{\dagger},k}+\nonumber\\
\tau^{c}_{u,k^{\dagger},k}+\tau^{t_b}_{u,k^{\dagger},k}+\tau^{d}_{u,k^{\dagger},k}\big)\Big]+\tau^m.
\end{align}
In (\ref{Sb_uk})\text{-}(\ref{kb}), substituting $k^{\prime\prime}=k^{\dagger}$ and $k^{\prime}=k$, if $\tau^b_{u,k^{\dagger},k}$ meets the threshold $T^b_{k^{\dagger},k}$, $w^b_{u,k^{\dagger},k}$ is feasible and $w^b_{u,k^{\dagger},k}\in \mathcal W^b_{u,k}$. Then, $\mathcal W^b_{u,k}\neq\varnothing$, $\mathbb{I}^b_{u,k}=1$. Before the merging stage of the $k$-th viewport frame, selecting the latest feasible background $w^b_{u,k^{\star},k}$. Otherwise, if $\mathbb{I}^f_{u,k}=0$ or $\mathbb{I}^b_{u,k}=0$, $\mathbb{I}_{u,k}$ is zero and ATW is utilized to generate the corresponding item.

\subsection{Optimization Problem}
In this paper, our goal is to optimize the control decisions and the MEC resource allocations for each user at each time slot to minimize the total cost while adhering to the MTP threshold. The optimization problem can be formulated as
\begin{align}
\min_{     
        \boldsymbol{z}^f_k, \boldsymbol{x}^b_k, \boldsymbol{z}^b_k,
        \boldsymbol{B}_k, \boldsymbol{F}^{\nu}_k}& 
        \lim_{K\rightarrow+\infty}\mathrel{\mspace{-2mu}}\frac{1}{K}\sum_{k=1}^{K} \left(\sum_{u\in \mathcal{U}}\left(\varkappa_{u,k}+\zeta\,\varepsilon_{u,k} \right)\right)\label{obj} \\
\text { s.t. }\quad\ &\lim_{K\rightarrow+\infty}\mathrel{\mspace{-1mu}}\frac{1}{K}\sum_{k=1}^{K}\left(\sum_{u\in\mathcal{U}}\left(1-\mathbb{I}_{u,k}\right)\right)\mathrel{\mspace{-3mu}}\leq\mathrel{\mspace{-3mu}}\epsilon,\tag{\ref{obj}{a}}\label{mtp constraint}\\
& \sum_{u\in\mathcal{U}}B_{u,k}\leq B,\tag{\ref{obj}{b}}\label{B_total constraint}\\
& \sum_{u{\setlength\arraycolsep{5pt}\in} \mathcal{U}} F^{\nu}_{u,k} \leq F^{\nu}.\tag{\ref{obj}{c}}\label{Fv_total constraint}
\end{align}
\label{p1}where $\zeta>0$ is a weighting parameter to balance the tradeoff between the sensor age and the VR device power consumption. $\boldsymbol{z}^f_k\triangleq(z^f_{u,k})_{u\in\mathcal{U}}$, $\boldsymbol{x}^b_k\triangleq(x^b_{u,k,l})_{u\in\mathcal{U}, l\in\mathcal{L}_k}$ and $\boldsymbol{z}^b_k\triangleq(z^b_{u,k,l})_{u\in\mathcal{U}, l\in\mathcal{L}_k}$ are the binary control decisions. The allocation of bandwidth resources at BS and GPU resources at MEC must satisfy $\boldsymbol{B}_k\succeq0$ and $\boldsymbol{F}^{\nu}_k\succeq0$, respectively. Constraint (\ref{mtp constraint}) aims to generate viewport frames through rendering rather than ATW. The probability of $\mathbb{I}_{u,k}=0$ in (\ref{mtp constraint}) is restricted to a relatively small average range $\epsilon\to0^{+}$, ensuring the reliability of end-to-end latency in meeting the MTP threshold. Constraints (\ref{B_total constraint}) and (\ref{Fv_total constraint}) represent that $\boldsymbol{B}_k$ and $\boldsymbol{F}^{\nu}_k$ at MEC are within the total bandwidth $B$ and total GPU resource $F^{\nu}$, respectively. 

\section{Design of AQM-CUP algorithm}\label{safe_RL_alg}
\begin{figure*}[t]
\centering{\includegraphics[width=0.85\textwidth]{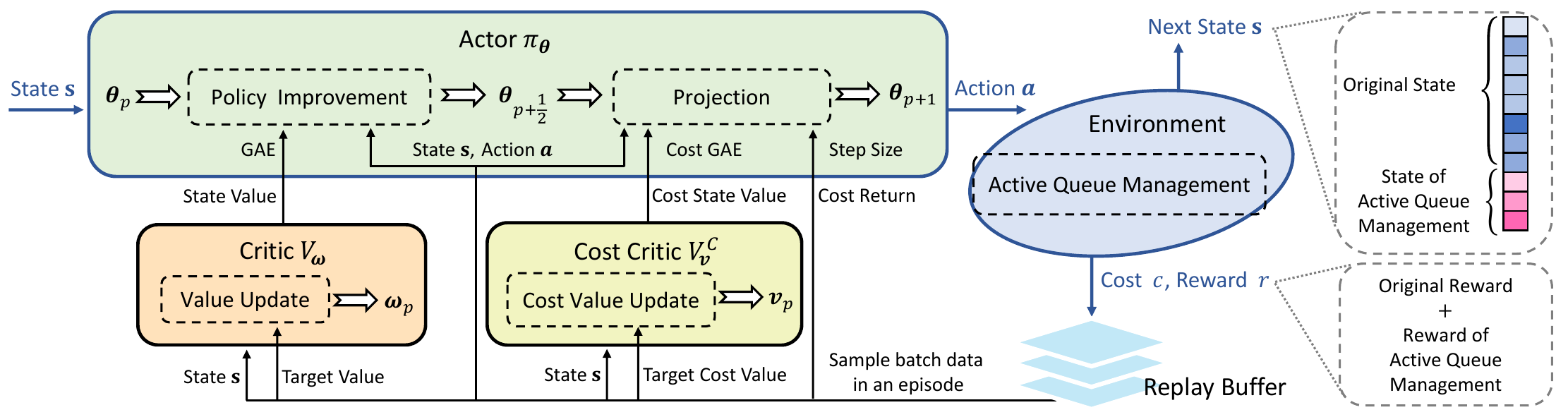}}
	\caption{Illustration of proposed AQM-CUP architecture.}
	\label{cup_aqm}
\end{figure*}

In this section, we formulate the optimization problem as a CMDP \cite{CMDP}. The intelligent agent in safe RL aims to maximize reward while adhering to safe constraints. To ensure theoretical safety, we use the CUP algorithm, a policy optimization method in safe RL. The  queue environment's state challenges for policy updates. During RL training, random actions can lead to queue congestion, invalidating subsequent actions. To address this, we integrate AQM into safe RL and design the AQM-CUP algorithm. The overall illustration of AQM-CUP architecture is shown in Fig. \ref{cup_aqm}.

\subsection{CMDP}
The CMDP is represented by a 4-tuple $\left(\mathcal{S}, \mathcal{A}, \mathcal{R}, \mathcal{C} \right)$, consisting of state space $\mathcal{S}$, action space $\mathcal{A}$, reward space $\mathcal{R}$, and cost space $\mathcal{C}$.

\subsubsection{State space} Each state is characterized by features extracted from environment observations. The BS acts as the agent, possessing comprehensive information about all mobile devices. Specifically, the state at the beginning of the $k$-th time slot, denoted as $\bm{s}_k \in \mathcal{S}$, encompasses flattened information from three components: mobile devices, requests from the latest joined viewport frame, and processing queues. The mobile devices component includes information $h_{u,k}g_{u,k}$ and the rendering capability $f_u^{\nu}$, where $u\in \mathcal{U}$. The requests component  from the latest joined viewport frame includes the foreground computation load $N^f_{u,k}$, the foreground data size $D^f_{u,k}$, the background computation load $N^b_{u,k,l}$, and the background data size $D^b_{u,k,l} = D^b$, where $u\in\mathcal{U}$, $l\in\mathcal{L}_k$. Denote $M$ as the length of each queue in the state, and the order set of queues is denoted as $\mathcal{M} = \{1,2,\cdots,M\}$. The processing queues component includes the rendering rendering load $N^{r_e,k}_{u,m}$, the awaiting data size $D^{r_e,k}_{u,m}$, and the remaining available MTP duration $\tau^{r_e,k}_{u,m}$ for the $m$-th order foreground or background tile in the MEC rendering queue $Q^{r_e}_u$; the awaiting data size $D^{c,k}_{u,m}$, the remaining available MTP duration $\tau^{c,k}_{u,m}$ for the $m$-th order background tile in the compressing queue $Q^{c}_u$; the awaiting data size $D^{t,k}_{u,m}$, the remaining available MTP duration $\tau^{t,k}_{u,m}$ for the $m$-th order foreground or background tile in the transmission queue $Q^{t}_u$; the awaiting data size $D^{d,k}_{u,m}$, the remaining available MTP duration $\tau^{d,k}_{u,m}$ for the $m$-th order background tile in the decompressing queue $Q^{d}_u$; the rest rendering load $N^{r_d,k}_{u,m}$, the remaining available MTP duration $\tau^{r_d,k}_{u,m}$ for the $m$-th order foreground or background tile in the device rendering queue $Q^{r_d}_u$, where $u\in\mathcal{U}, m\in\mathcal{M}$.

\subsubsection{Action space} Let $\bm{a}_k \in \mathcal{A}$ be the action at the $k$-th time slot, comprises the flattened optimization variables $\boldsymbol{z}^f$, $\boldsymbol{x}^b$, $\boldsymbol{z}^b$, $\boldsymbol{B}_k$, and $\boldsymbol{F}^{\nu}_k$.


\subsubsection{Instantaneous reward and cost} Define $r_{k+1}$ and $c_{k+1}$ as the instantaneous reward and cost, respectively, obtained after performing the action $\bm{a}_k$ under the state $\bm{s}_k$. Both $r_{k+1}\in \mathcal{R}$ and $c_{k+1}\in\mathcal{C}$ are scalars in $\mathbb{R}$. According to (\ref{tau_qoe}), (\ref{energy_qoe}), and (\ref{obj}), the instantaneous reward is expressed as
\begin{equation}
r_{k+1}= -\sum_{u\in\mathcal{U}}\left(\varkappa_{u,k}+\zeta\,\varepsilon_{u,k} \right).
\end{equation}
Based on (\ref{mtp constraint}), the instantaneous cost is given by
\begin{equation}\label{total_c}
c_{k+1}= \sum_{u\in\mathcal{U}}\left(1-\mathbb{I}_{u,k}\right).
\end{equation}

\subsection{Queue Environment}
Based on the queue environment $Q^{\Upsilon}_u$ of user $u$, $\Upsilon \in \{ r_e, c, t, d, r_d\}, u\in \mathcal{U}$, given $\bm{s}_k$ and $\bm{a}_k$, the next state $\bm{s}_{k+1}$, the instantaneous reward $r_{k+1}$ and the instantaneous cost $c_{k+1}$ are obtained as follows. 
\subsubsection{Latest tiles enter rendering queues} At the beginning of the $k$-th time slot, the awaiting rendering tiles are allocated to $Q_u^{r_e}$ or $Q_u^{r_d}$ according to the control decisions in $\bm{a}_k$ of user $u$, i.e., $z^f_{u,k}$, $x^b_{u,k,l}$, $z^b_{u,k,l}$, $l\in\mathcal{L}_k$. The total duration of each tile is initialized to zero.
\subsubsection{Dequeue and enqueue based on timing sequence} 
Take $Q^{t}_u$ as an example. First, record the current timestamp and the remaining latency of this time slot. Next, check whether there are pending tiles in $Q^{t}_u$. If $Q^{t}_u$ is non-empty, iteratively transmit the first data at the front of $Q^{t}_u$ according to the following two steps: (\romannumeral1) Check if the entry timestamp, $I^{t_{\varphi}}_u, \varphi\in\{f,b\}$, falls within the range of this time slot: If so, there is a waiting delay before transmission. Update the current timestamp to $I^{t_{\varphi}}_u$ and subtract the waiting delay from the remaining time. If the data arrives early, start transmission immediately. If the data arrives late, there is no data ready for transmission during this slot, so exit the loop. (\romannumeral2) Transmit the data according to the allocated resource $B_{u,k}$. Compare the required delay of the first data in $Q^{t}_u$ with the remaining time of this time slot.If the remaining time is greater than or equal to the required transmission delay, the tile dequeues during this time slot, i.e., the entire tile arrives at the mobile device, and the total duration of both the dequeued tile and each tile in $Q^{t}_u$ increases. Then, the mobile device checks the type of received tile, adding foreground tiles to the awaiting merging pool and background tile to $Q^d_u$. If the remaining time of this time slot is greater than the required transmission delay, continue looping; if it is equal, exit the loop. If the remaining time is less than the required transmission delay, a portion of the data is transmitted during this loop, but not entirely. The data size of the front pending tile in $Q^t_u$ decreases by the transmitted portion. Increase the total duration for each tile in $Q^t_u$, and exit the loop. The queue environments for $Q^{r_e}, Q^{c}, Q^d, Q^{r_d}$ follow a similar pattern and will not be described repeatedly.
\subsubsection{Generate frame and get results} In the current awaiting merging pool, check if there are foreground and background tiles of the $k$-th frame. If present, assess their feasibility, i.e., whether their total duration plus $\tau^m$ exceeds the corresponding MTP thresholds $T^f$ and $T^b_{k^{\dagger},k}$. Finally, $\bm{s}_{k+1}$, $r_{k+1}$, $c_{k+1}$ can be obtained. To clean up timeout content in the awaiting merging pool, only tiles beyond the $k$-th frame are retained. 

\subsection{AQM Architecture}

We use AQM to remove tiles that have already fallen below the lower bound of the required time from the queue. Denote the remaining MTP threshold of $m$-th order tile in queue $Q^{\Upsilon}_u$ at the $k$-th time slot as $\tau^{\text{rest}_k, \Upsilon}_{u,m}$, where $m\in\mathcal{M}$, $\Upsilon \in \{ r_e, c, t, d, r_d\}$. Drop the tiles in $Q^{r_e}_u$ with $\tau^{\text{rest}_k,r_e}_{u,m}\leq\tau^m$ for the foreground tile, and $\tau^{\text{rest}_k,r_e}_{u,m}\leq(\tau^m+\Delta^c+\Delta^d)$ for the background tile. Drop the tiles in $Q^{c}_u$ with $\tau^{\text{rest}_k,c}_{u,m}\leq(\tau^m+\Delta^d)$. Drop the tiles in $Q^{t}_u$ with $\tau^{\text{rest}_k,t}_{u,m}\leq\tau^m$ for the foreground tile, and $\tau^{\text{rest}_k,t}_{u,m}\leq(\tau^m+\Delta^d)$ for the background tile. Drop the tiles in $Q^{d}_u$ with $\tau^{\text{rest}_k,d}_{u,m}\leq\tau^m$. Drop the tiles in $Q^{r_d}_u$ with $\tau^{\text{rest}_k,r_d}_{u,m}\leq\tau^m$. 

The number of discarded tiles in queue $Q^{\Upsilon}_u$ at the $k$-th time slot for user $u$ is denoted as $\Lambda_{u,k}^{\Upsilon}, u\in\mathcal{U}, \Upsilon\in\{ r_e, c, t, d, r_d\}$. The state of AQM is $\Lambda_{u,k}^{\Upsilon}$ with a dimension of $5 U$. The total state is formed by concatenating the AQM state with the original $\bm{s}_k$. Since the discarded tiles have consumed resources, the AQM reward is defined to penalize the dropping tiles, calculated as $-10^{-4}\sum_{u, \Upsilon}\Lambda_{u,k}^{\Upsilon}$. The total instantaneous reward is the sum of the original reward and the AQM reward,
\begin{equation}\label{total_r}
r_{k+1}= -\sum_{u\in\mathcal{U}}\left(\varkappa_{u,k}+\zeta\,\varepsilon_{u,k} \right)-10^{-4}\sum_{u, \Upsilon}\Lambda_{u,k}^{\Upsilon}.
\end{equation}
 
\subsection{Definitions in safe RL}
The goal of CMDP is to search the optimal policy $\pi_{\star}$:
\begin{align}
\pi_{\star}=\arg &\max _{\pi_{\boldsymbol{\theta}} \in \Pi_{\boldsymbol{\theta}}} J\left(\pi_{\boldsymbol{\theta}}\right)=\mathbb{E}_{\pi_{\boldsymbol{\theta}}}\left[\sum_{k=0}^{\infty} \gamma^k r\left(\bm{s}_k, \bm{a}_k\right)\right] \\
&\text { s.t. }J^{C}\left(\pi_{\boldsymbol{\theta}}\right) {\leq} \epsilon
. \label{JC_constaint}
\end{align}
where $J$ is the expected return, $\gamma$ is a discount rate, and $J^{C}$ is cost-return. The estimated cost-return on the $i$-th episode is
\begin{equation}\label{JC}
\hat{J}_{i}^C = \sum_{k=0}^{K} \gamma^k c_{i,k+1}, 
\end{equation} 
where $i\in\{1,\cdots,H\}$ and $H$ is the track number, $K$ is the trajectory horizon. 


Generalized advantage estimator (GAE) is used to measure the effectivity of action under a certain state. The estimated GAE for the $i$-th episode is given by\cite{GAE}:
\begin{equation}\label{A}
\hat{A}_{i, k}=\sum_{j=k}^K(\gamma \lambda)^{j-k} \delta_{i, j},
\end{equation}
where $\lambda \in[0,1]$, and TD error is denoted by 
\begin{equation}\label{delta}
\delta_{i, k}=r_{i, k}+\gamma V_{\bm{\omega}_{p}}\left(\bm{s}_{i, k}\right)-V_{\bm{\omega}_{p}}\left(\bm{s}_{i, k-1}\right),
\end{equation}
where $V_{\bm{\omega}_{p}}$ is an estimator of value function on the $p$-th iteration. Similarly, the cost GAE is given by
\begin{equation}\label{AC}
\hat{A}_{i, k}^C=\sum_{j=k}^K(\gamma \lambda)^{j-k} \delta_{i, j}^C,
\end{equation}
where the cost TD error is 
\begin{equation}\label{deltaC}
\delta_{i, k}^C=c_{i, k}+\gamma V_{\bm{v}_{p}}^C\left(\bm{s}_{i, k}\right)-V_{\bm{v}_{p}}^C\left(\bm{s}_{i, k-1}\right),
\end{equation}
$V_{\bm{v}_{p}}^C$ is an estimator of cost function $c$ on the $p$-th iteration.

\subsection{Preliminary of CUP}
With one-step update in policy gradients \cite{CPO}, CMDP can be transformed as follows
\begin{align}
\pi_{\bm_{p+1}}=\arg &\max _{\pi \in \Pi_{\boldsymbol{\theta}}} J\left(\pi\right) \label{one_step_obj}\\
\text { s.t. }&J^{C}\left(\pi\right) {\leq} \epsilon, \label{JC_constraint2}\\
& \mathrm {KL} (\pi_p, \pi){\leq} \epsilon^{\prime}\label{KL_constraint}, 
\end{align}
where $\mathrm {KL} (\pi_p, \pi)$ denotes the KL-divergence with respect to $\pi_p$ and $\pi$. (\ref{KL_constraint}) implies the distribution of update policy output is close to the distribution of original policy output. 

For CUP algorithm\cite{CUP}, each update contains two sub-steps: policy improvement and projection.
\subsubsection{Policy improvement}
According to PPO\cite{PPO}, for mini-
epoch number $\hat{p}=\{1,2,\dots,\Gamma\}$, sample mini-batch with size $\hat{H}$ of trajectory under policy $\pi_{\bm{\theta}_p}$, i.e., $\cup_{i=1}^{\hat{H}} \cup_{k=0}^{K} \{(\bm{s}_{i,k}, \bm{a}_{i,k}, r_{i,k+1}, c_{i,k+1})\}$. KL divergence is replaced by a clip implementation, then policy improvement is denoted by,
\begin{align}
\bm{\theta}_{p+\frac{1}{2}} = &\arg\max_{\bm{\theta}}\Bigg\{\sum_{i=1}^{\hat{H}}\sum_{k=0}^{K}\min \bigg\{ \frac{\pi_{\boldsymbol{\theta}}\left(\bm{a}_{i,k}\mid\bm{s}_{i,k}\right)}{\pi_{\boldsymbol{\theta}_p}\left(\bm{a}_{i,k}\mid\bm{s}_{i,k}\right)} \hat{A}_{i,k}, \nonumber\\&\mathrm{clip}\left(\frac{\pi_{\boldsymbol{\theta}}\left(\bm{a}_{i,k}\mid\bm{s}_{i,k}\right)}{\pi_{\boldsymbol{\theta}_p}\left(\bm{a}_{i,k}\mid\bm{s}_{i,k}\right)}, 1-\chi,1+\chi\right)\hat{A}_{i,k}\bigg\}\Bigg\},\label{policy improvement}
\end{align}
where $\chi$ is a positive decimal value.
\subsubsection{Projection}
When $\mathbb{E}_{\boldsymbol{s} \sim \Omega_{\pi_{\boldsymbol{\theta}_p}(\cdot)}}\!\!\left[\mathrm{KL}\!\left(\pi_{\boldsymbol{\theta}_p}(\cdot \mid s), \pi_{\boldsymbol{\theta}}(\cdot \mid s)\right)\!\right]\!\!\to\!0$ 
and the sampled trajectory under policy $\pi_{\bm{\theta}_p}$ is substituted, step-size $v$ is updated as follows
\begin{align}
v_{p+1} = \langle\, v_{p} + \eta(\hat{J}^C_p - \epsilon)\,\rangle^+,\label{lagrange_multip}
\end{align}
where $\eta$ is a positive constant. Projection is denoted by
\begin{align}
\bm{\theta}_{p+1}= \arg \min_{\bm{\theta}}   &\sum_{i=1}^{\hat{H}}\sum_{k=0}^{K} \Big\{  \mathrm {KL} (\pi_{\bm{\theta}_{p+\frac{1}{2}}}(\cdot |\bm{s}_{i,k}), \pi_{\bm{\theta}}(\cdot |\bm{s}_{i,k}))\nonumber\\ 
&+ v_{p} \frac{1-\gamma \lambda}{1-\gamma}\frac{\pi_{\bm{\theta}}(\bm{a}_{i,k}|\bm{s}_{i,k})}{\pi_{\bm{\theta}_{p}}(\bm{a}_{i,k}|\bm{s}_{i,k})} \hat{A}_{i,k}^C \Big\}. \label{projection}
\end{align}

In general, the detailed information of the proposed AQM-CUP algorithm is presented in Algorithm \ref{AQM_CUP_alg}.

\subsection{Complexity Analysis}\label{complexity}
The AQM-CUP algorithm of network training contains policy network training in policy improvement and projection, value network training, cost value network training. Denote the dimension of state $\bm{s}_k$ and action $\bm{a}_k$ as $(X_{\bm{s}}, X_{\bm{a}})$. Denote the number of hidden layers in policy network, value network and cost value network as $(Y_{\bm{\theta}}, Y_{\bm{\omega}}, Y_{\bm{v}})$. Denote the number of neurons for hidden layer $e$ in policy network, value network and cost value network as $(Z_{\bm{\theta}}^{e}, Z_{\bm{\omega}}^{e}, Z_{\bm{v}}^{e})$. The computational complexity of single step for the policy network in policy improvement and projection is $\mathcal{O}(2(X_{\bm{s}}Z_{\bm{\theta}}^{1}+\sum_{e=1}^{Y_{\bm{\theta}}-1}Z_{\bm{\theta}}^{e}Z_{\bm{\theta}}^{e+1}+X_{\bm{a}}Z_{\bm{\theta}}^{Y_{\bm{\theta}}}))$. The computational complexities of single step for the value network and cost value network are $\mathcal{O}(X_{\bm{s}}Z_{\bm{\omega}}^{1}+\sum_{e=1}^{Y_{\bm{\omega}}-1}Z_{\bm{\omega}}^{e}Z_{\bm{\omega}}^{e+1}+Z_{\bm{\omega}}^{Y_{\bm{\omega}}})$ and $\mathcal{O}(X_{\bm{s}}Z_{\bm{v}}^{1}+\sum_{e=1}^{Y_{\bm{v}}-1}Z_{\bm{v}}^{e}Z_{\bm{v}}^{e+1}+Z_{\bm{v}}^{Y_{\bm{v}}})$. Therefore, the total computational complexity of the proposed algorithm is $\mathcal{O}(\Gamma \hat{H} (K+1) (2(X_{\bm{s}}Z_{\bm{\theta}}^{1}+\sum_{e=1}^{Y_{\bm{\theta}}-1}Z_{\bm{\theta}}^{e}Z_{\bm{\theta}}^{e+1}+X_{\bm{a}}Z_{\bm{\theta}}^{Y_{\bm{\theta}}})+
(X_{\bm{s}}Z_{\bm{\omega}}^{1}+\sum_{e=1}^{Y_{\bm{\omega}}-1}Z_{\bm{\omega}}^{e}Z_{\bm{\omega}}^{e+1}+Z_{\bm{\omega}}^{Y_{\bm{\omega}}})+
(X_{\bm{s}}Z_{\bm{v}}^{1}+\sum_{e=1}^{Y_{\bm{v}}-1}Z_{\bm{v}}^{e}Z_{\bm{v}}^{e+1}+Z_{\bm{v}}^{Y_{\bm{v}}})))$.
\begin{algorithm}[t]
\caption{AQM-CUP}
\begin{algorithmic}[1]\label{AQM_CUP_alg}
\STATE \textbf{Initialize:} policy network $\bm{\theta}_0$, value network $\bm{\omega}_0$, cost value network $\bm{v}_0$, step-size $v_0$;
\STATE \textbf{Parameters:} track number $H$, trajectory horizon $K$, mini-epoch number $\Gamma$, mini-batch size $\hat{H}$, discount rate $\gamma$, discount factor $\lambda$, positive constant $\eta$;

\FOR{${p} = 0, 1, 2, \dots$}
    \FOR{$i=1,2,\dots, H$, and $k = 0,1,\dots, K$} 
    \STATE Utilize AQM to drop timeout tiles in queues;
    \STATE Get state of AQM to form the total state $\bm{s}_{i,k}$, \\$\quad$
    reward of AQM;
    \STATE Choose $\bm{a}_{i,k}$ based on $\bm{s}_{i,k}$ according to $\pi_{\bm{\theta}_p}$;
    \STATE Update queues based on environment;
    \STATE Get instantaneous total reward $r_{i,k+1}$: (\ref{total_r}),\\$\quad$
     instantaneous total cost $c_{i,k+1}$: (\ref{total_c});   
    \STATE Store the sample $(\bm{s}_{i,k}, \bm{a}_{i,k}, r_{i,k+1}, c_{i,k+1})$;
    \ENDFOR
    \STATE Estimate cost-return $\hat{J}_{i}^C$: (\ref{JC}),\\ $\quad$ average cost-return $\hat{J}_{p}^C$: $\hat{J}_{p}^C=\frac{1}{H} \sum_{i=1}^{H} \hat{J}_{i}^C$;
    \STATE Compute TD errors $\cup_{i=1}^H \cup_{k=0}^K\{\delta_{i, k}\}$: (\ref{delta}),\\ $\quad$ cost TD errors $\cup_{i=1}^H \cup_{k=0}^K\{\delta_{i, k}^C\}$: (\ref{deltaC});
\STATE Compute GAE $\hat{A}_{i, k}$: (\ref{A}),\\$\quad$ cost GAE $\hat{A}^C_{i, k}$: (\ref{AC});
\STATE Compute target value function $V_{i, k}^{\text {target }}$, \\
$\quad$ target cost value function $V_{i, k}^{\text {target}, C}$: 
\begin{align}
    V_{i, k}^{\text {target }}=\hat{A}_{i, k}+V_{\boldsymbol{\omega}_{p}}\left(\bm{s}_{i, k}\right), \, \nonumber\\
    V_{i, k}^{\text {target}, C}=\hat{A}_{i, k}^C+V_{\boldsymbol{v}_{p}}^C\left(\bm{s}_{i, k}\right);\nonumber
\end{align}
\STATE Store data $\mathcal{D}_{p}$: \\\ \ $\cup_{i=1}^{H} \cup_{k=0}^{K}\left\{({\bm{a}_{i,k}, \bm{s}_{i,k}, \hat{A}_{i,k}, \hat{A}^{C}_{i,k}, V^{\text{target}}_{i,k}, V^{\text{target}, C}_{i,k}})\right\}$;
\FOR {$\hat{p}=1,2,\dots,\Gamma$, sample mini-batch from $\mathcal{D}_{p}$}
    \STATE Get $\bm{\theta}_{p+\frac{1}{2}}$ by policy improvement: (\ref{policy improvement});
\ENDFOR

\STATE Get step-size $v_{p+1}$: (\ref{lagrange_multip});

\FOR {$\hat{p}=1,2,\dots,\Gamma$, sample mini-batch from $\mathcal{D}_{p}$}
    \STATE 
    Get $\bm{\theta}_{p+1}$ by projection: (\ref{projection});
\ENDFOR

\FOR {$\hat{p}=1,2,\dots,\Gamma$, sample mini-batch from $\mathcal{D}_{p}$}
    \STATE Update value network, cost value network: \begin{align}
    \quad \bm{\omega}_{p+1}=\arg \min _{\bm{\omega}} \sum_{i=1}^{\hat{H}}\sum_{k=0}^{K}\left(V_{\bm{\omega}}\left(\bm{s}_{i,k}\right)-V_{i,k}^{\text{target}}\right)^2,\,\nonumber\\
    \bm{v}_{p+1}=\arg \min _{\bm{v}} \sum_{i=1}^{\hat{H}}\sum_{k=0}^{K}\left(V_{\bm{v}}^C\left(\bm{s}_{i,k}\right)-V_{i,k}^{\text{target}, C}\right)^2 ;\nonumber\end{align}
\ENDFOR
\ENDFOR
\end{algorithmic}
\end{algorithm}
 
\section{Simulation Results}\label{simulation_results}
In this section, we conduct comprehensive experiments to evaluate the performance of the proposed framework and algorithm. We consider multiple users with HMDs are uniformly distributed in a $20\mathrm{m}\times20\mathrm{m}$ room. The mmWave BS is located at the center of the room and we consider the LOS and NLOS probability for the indoor scenario based on 3GPP\cite{3GPP_mmwave}. The shadowing fading loss for LOS and NLOS cases are respectively $\ell^{\operatorname{LOS}_{\varsigma}}=3$ and $\ell^{\operatorname{NLOS}_{\varsigma}}=8.03$ (in $\mathrm{dB}$). We set $f^o=28\,\mathrm{GHz}$, $\phi=30^{\circ}$, $g^\mu=10\,\mathrm{dB}$, $g^{\kappa}=-10\,\mathrm{dB}$, $P=30\,\mathrm{dBm}$, $N_0=-147\,\mathrm{dBm/Hz}$.

A display with a resolution of $2064 \times 2208\,\mathrm{pixels}$ is considered \cite{meta_quest3}, i.e., $n^{p_b} = 2064 \times 2208\,\mathrm{pixels}$. Each pixel has 24 bits of RGB data and 16 bits of depth data, totaling $40\,\mathrm{bits}$, i.e., $\varrho= 40\,\mathrm{bits/pixel}$. The raw background data size of each tile is $D^b = 40 \times 2064 \times 2208 \approx 182\,\mathrm{Mbits}$. The background tiles are encoded using H.264, achieving a compression ratio of $\alpha = 1.6\%$. All foreground objects account for $0\%\sim50\%$ of the total visual field pixels, i.e., $n^{p_f}_{u,k}/n^{p_b}\in [0, 0.5]$. Foreground objects and background environment are selected from Sketchfab\cite{Sketchfab}. To simulate multi-user interactive VR, we record $H=1000$ tracks of foreground and background parameters for $U=5$ users over $K = 300$ time slots \cite{Q-vr} to construct the training dataset. In the collected tracks, $n^{v_f}_{u,k}=1\sim40\,\mathrm{K}$, $n^{v_{b}}_{u,k^{\dagger},k}=10\sim20\,\mathrm{K}$, $c^{v_f}_{u,k} = 100 \sim600\,\mathrm{cycles/pixel}$, $c^{v_{b}}_{u,k^{\dagger},k} = 100 \sim200\,\mathrm{cycles/pixel}$, $c^{p_f}_{u,k} = 5 \sim 50\,\mathrm{cycles/pixel}$, $c^{p_{b}}_{u,k^{\dagger},k} = 5\sim20\,\mathrm{cycles/pixel}$. The MTP latency threshold is $T^{\text{MTP}}=20\,\mathrm{ms}$. The frame rate is set to $F = 100\,\mathrm{FPS}$, and $\tau=10\,\mathrm{ms}$. We set $L=5$, $T=500\,\mathrm{ms}$, $\beta=10^{-25}\,\mathrm{s\cdot J/cycle}$, $\varepsilon^d =10\,\mathrm{J/s}$, $\tau^m=2\,\mathrm{ms}$, $\Delta^c=5\,\mathrm{ms}$, $\Delta^d=8\,\mathrm{ms}$, $f^{\nu}_{u}=3\,\mathrm{GHz}$, $B=500\,\mathrm{MHz}$, $F^{\nu}=70\,\mathrm{GHz}$. The default parameters used in the algorithm are as follows, 
$\zeta=0.1$, $\epsilon=0$, $\Gamma=30$, $\hat{H}=200$, $\eta=0.01$, $\chi=0.2$, $v_0=0$, $\gamma=0.99$, $\lambda=0.95$.

\begin{figure}[t] 
\centering
\subfigure[Return in each episode.]{
\begin{minipage}[t]{0.45\linewidth}
\centering
\includegraphics[height=2.94cm, keepaspectratio]{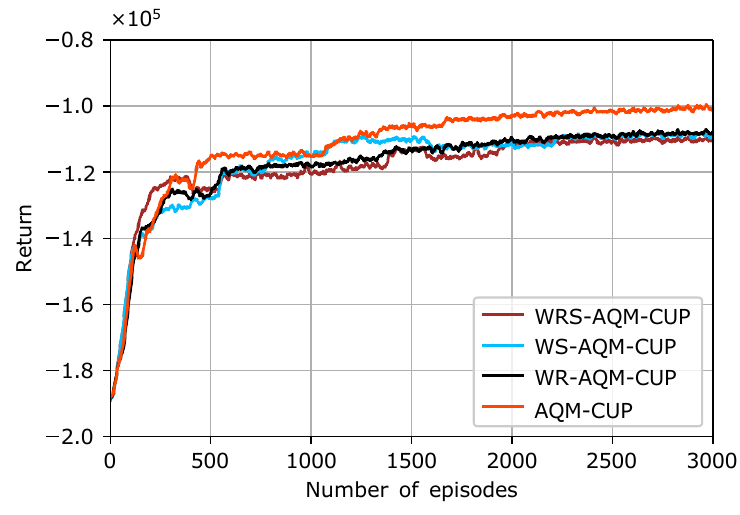}
\label{training_performance_reward}
\end{minipage}
}
\hspace{0.15cm}
\subfigure[Cost return in each episode.]{
\begin{minipage}[t]{0.45\linewidth}
\centering
\includegraphics[height=2.94cm, keepaspectratio]{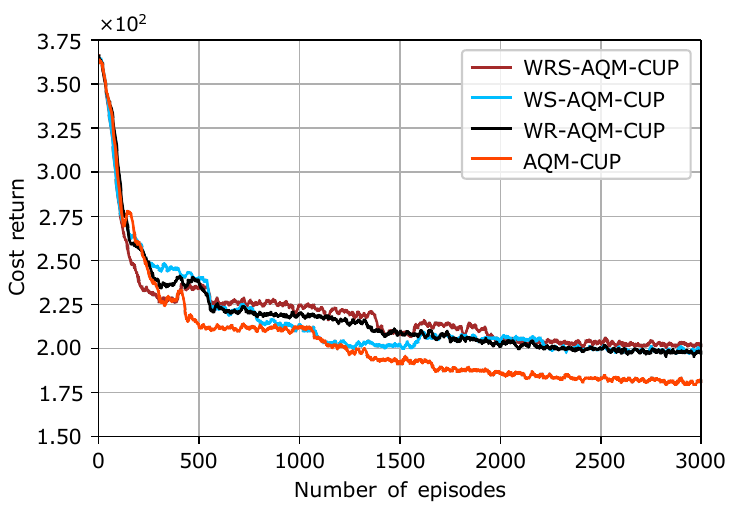}
\label{training_performance_cost}
\end{minipage}
}
\subfigure[Total age of sensor information in each episode.]{
\begin{minipage}[t]{0.45\linewidth}
\centering
\includegraphics[height=2.9cm, keepaspectratio]{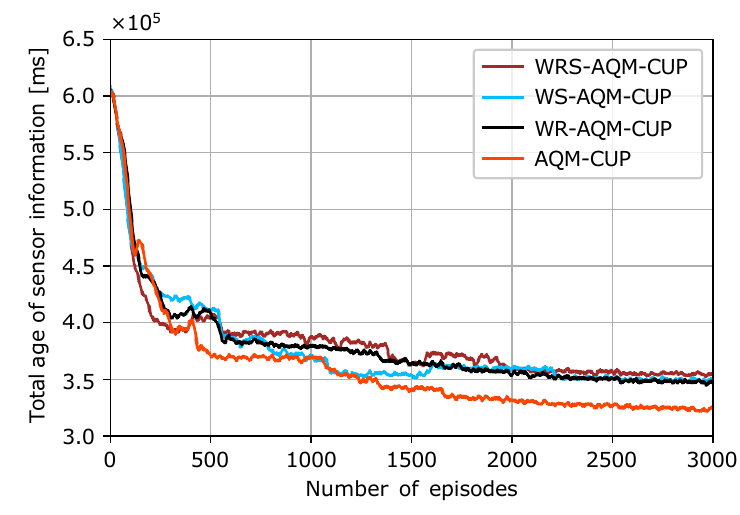}
\label{training_performance_age}
\end{minipage}
}
\hspace{0.15cm}
\subfigure[Total power consumption of devices in each episode.]{
\begin{minipage}[t]{0.45\linewidth}
\centering
\includegraphics[height=2.9cm, keepaspectratio]{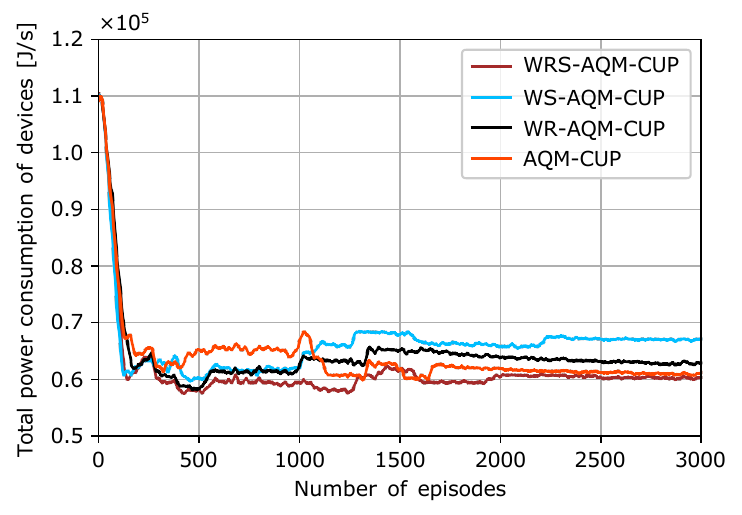}
\label{training_performance_energy}
\end{minipage}
}
\caption{Ablation study in the training performance of the AQM-CUP with $B=500\,\mathrm{MHz}$, $F^{\nu}=50\,\mathrm{GHz}$, $\zeta = 0.1$.}
\label{training_performance}
\end{figure}

The ablation study of the training performance for the AQM-CUP algorithm is shown in Fig. \ref{training_performance}. Three baselines are provided: (\romannumeral1) WS-AQM-CUP, the AQM-CUP algorithm without considering the state of AQM for the number of discarded tiles in queues; (\romannumeral2) WR-AQM-CUP, which excludes the reward of AQM for dropping tiles in queues; (\romannumeral3) WRS-AQM-CUP, which lacks both the state and reward of AQM, representing the benchmark with the most information loss. As depicted in Fig. \ref{training_performance_reward} and Fig. \ref{training_performance_cost}, the proposed AQM-CUP converges to better results. Fig. \ref{training_performance_age} illustrates the effectiveness of AQM-CUP in reducing the age of sensor information. In Fig. \ref{training_performance_energy}, AQM-CUP achieves the second-lowest power consumption for devices, following WRS-AQM-CUP. The suboptimal power efficiency of AQM-CUP is due to the low weight of power efficiency, e.g., $\zeta = 0.1$.

\begin{figure}[t] 
\centering
\subfigure[Return in each episode.]{
\begin{minipage}[t]{0.45\linewidth}
\centering
\includegraphics[height=2.94cm, keepaspectratio]{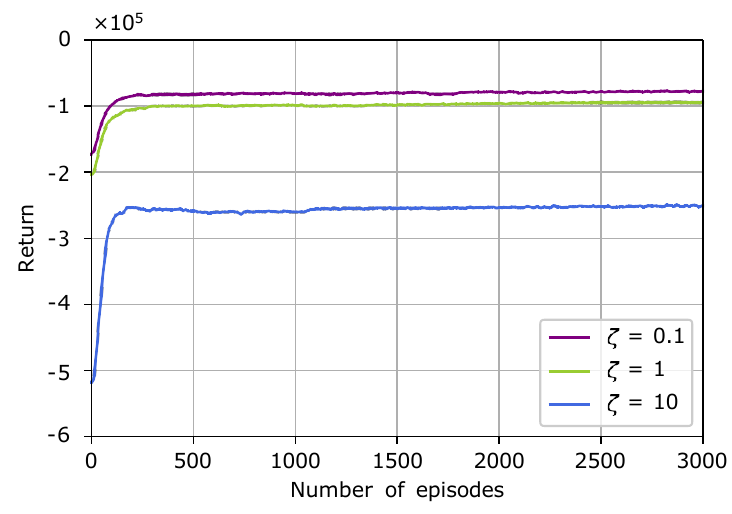}
\label{training_performance_zeta_reward}
\end{minipage}
}
\hspace{0.15cm}
\subfigure[Cost return in each episode.]{
\begin{minipage}[t]{0.45\linewidth}
\centering
\includegraphics[height=2.94cm, keepaspectratio]{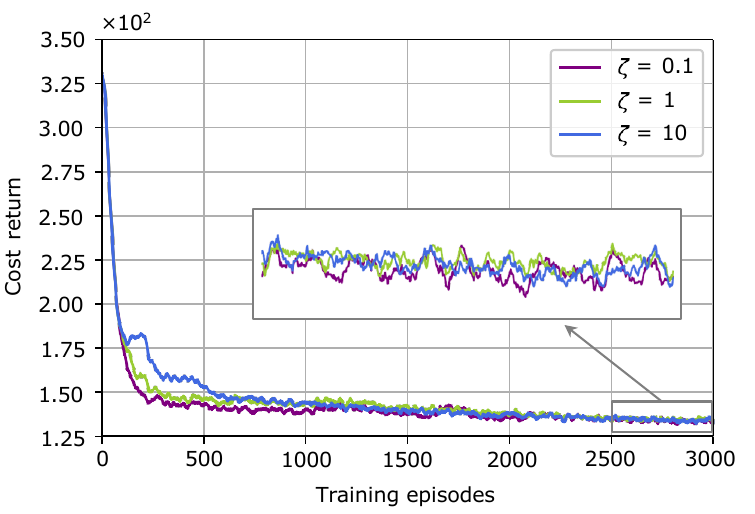}
\label{training_performance_zeta_cost}
\end{minipage}
}
\subfigure[Total age of sensor information in each episode.]{
\begin{minipage}[t]{0.45\linewidth}
\centering
\includegraphics[height=2.9cm, keepaspectratio]{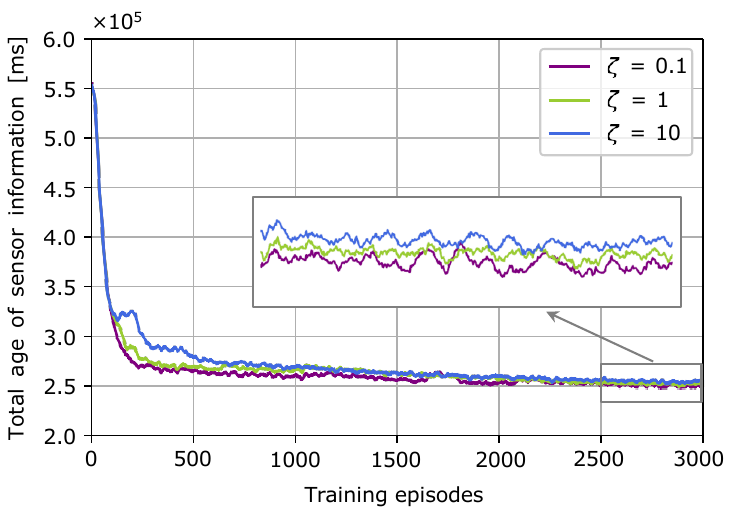}
\label{training_performance_zeta_age}
\end{minipage}
}
\hspace{0.15cm}
\subfigure[Total power consumption of devices in each episode.]{
\begin{minipage}[t]{0.45\linewidth}
\centering
\includegraphics[height=2.9cm, keepaspectratio]{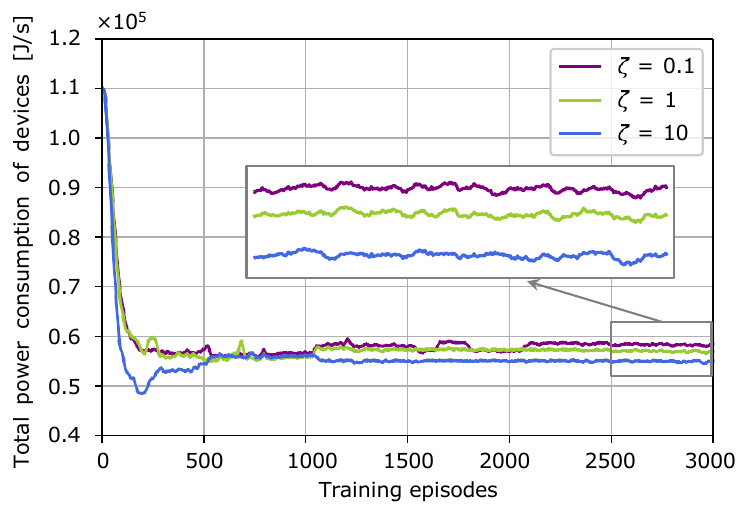}
\label{training_performance_zeta_energy}
\end{minipage}
}
\caption{Trade off parameter $\zeta$ in the training performance of the AQM-CUP with $B=800\,\mathrm{MHz}$, $F^{\nu}=90\,\mathrm{GHz}$.}
\label{training_performance_zeta}
\end{figure}

As depicted in Fig. \ref{training_performance_zeta}, we examine the impact of $\zeta$ on the training performance of the AQM-CUP algorithm. Fig. \ref{training_performance_zeta_reward} indicates that the return decreases as $\zeta$ increases. This is because the increase in $\zeta$ amplifies the power efficiency term in the reward (\ref{total_r}). From Fig. \ref{training_performance_zeta_cost}, we see that the training performance of cost return remains nearly constant across different $\zeta$ values. This is because the proposed AQM-CUP algorithm prioritizes the cost constraint over the metrics in the reward. In Fig. \ref{training_performance_zeta_age}, the metric of sensor information age slightly deteriorates as $\zeta$ increases, indicating that the increased weight on energy consumption decreases the importance of sensor information age. Furthermore, this change is inconspicuous because an appreciable penalty term $T$ is imposed in (\ref{tau_qoe}) when the MTP latency exceeds the threshold, making the freshness of prediction information less critical. Fig. \ref{training_performance_zeta_energy} shows that $\zeta$ significantly affects the metric of device power consumption, with the metric improving as $\zeta$ increases.

\begin{figure}[t] 
\centering
\subfigure[Average age of sensor information per user per time slot.]{
\begin{minipage}[t]{0.45\linewidth}
\centering
\includegraphics[height=2.94cm, keepaspectratio]{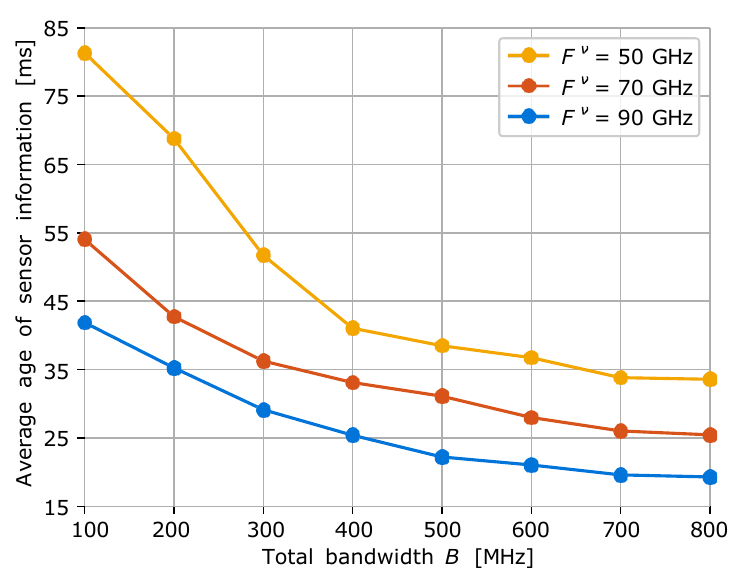}
\label{test_BF_age}
\end{minipage}
}
\hspace{0.15cm}
\subfigure[Average device power consumption per user per time slot.]{
\begin{minipage}[t]{0.45\linewidth}
\centering
\includegraphics[height=2.94cm, keepaspectratio]{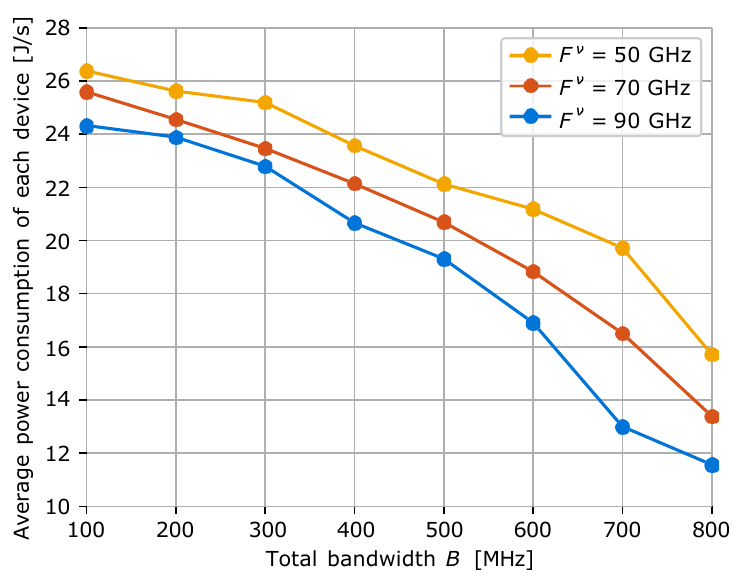}
\label{test_BF_energy}
\end{minipage}
}
\caption{Performance of the AQM-CUP vs. total bandwidth with different total GPU resources, $\zeta=0.1$.}
\label{test_BF}
\end{figure}

\begin{figure}[t] 
\centering
\subfigure[Average age of sensor information per user per time slot.]{
\begin{minipage}[t]{0.45\linewidth}
\centering
\includegraphics[height=2.94cm, keepaspectratio]{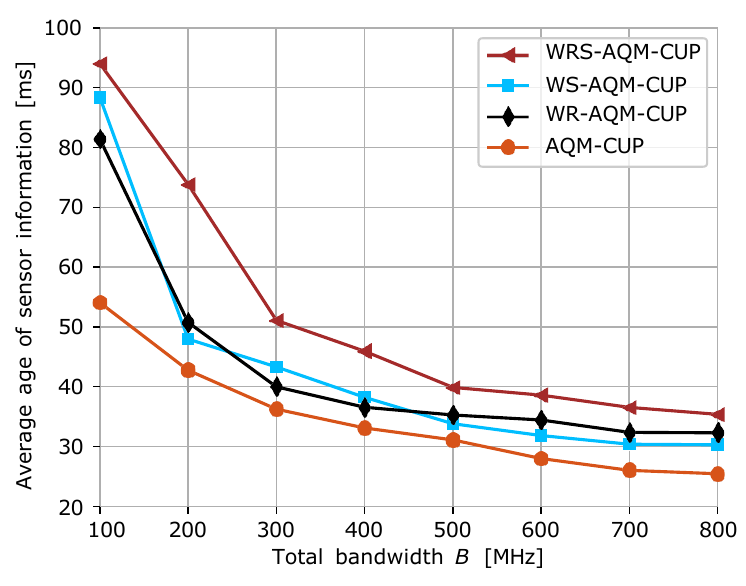}
\label{test_algorithm_age}
\end{minipage}
}
\hspace{0.15cm}
\subfigure[Average device power consumption per user per time slot.]{
\begin{minipage}[t]{0.45\linewidth}
\centering
\includegraphics[height=2.94cm, keepaspectratio]{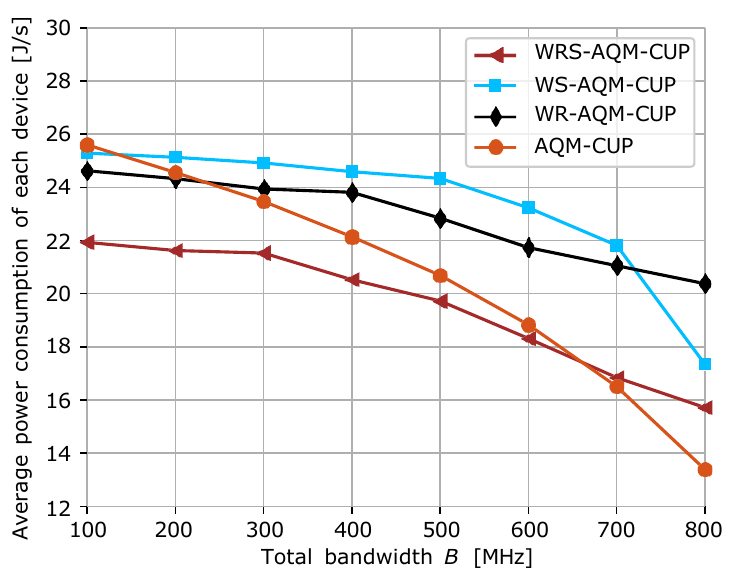}
\label{test_algorithm_energy}
\end{minipage}
}
\caption{Performance of different algorithms vs. total bandwidth with $F^{\nu}=70\,\mathrm{GHz}$, $\zeta=0.1$. }
\label{test_algorithm}
\end{figure}

\begin{figure}[t] 
\centering
\subfigure[Average age of sensor information per user per time slot.]{
\begin{minipage}[t]{0.45\linewidth}
\centering
\includegraphics[height=2.94cm, keepaspectratio]{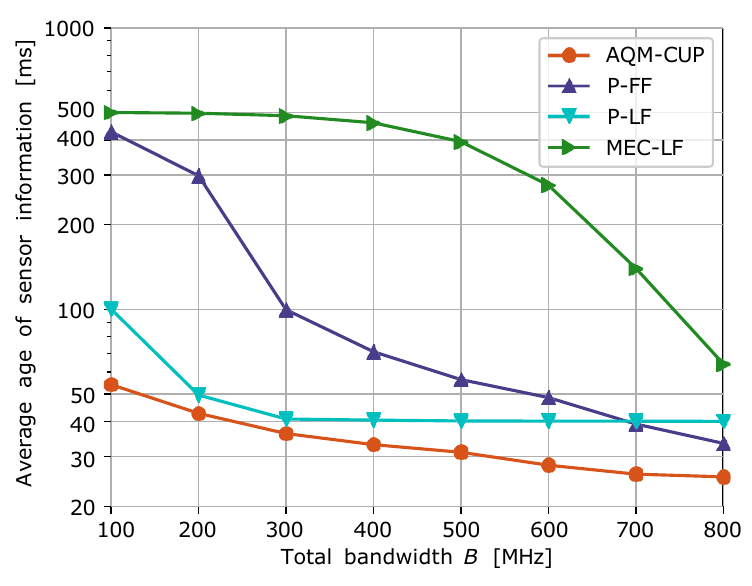}
\label{test_time_invariant_method_age}
\end{minipage}
}
\hspace{0.15cm}
\subfigure[Average device power consumption per user per time slot.]{
\begin{minipage}[t]{0.45\linewidth}
\centering
\includegraphics[height=2.94cm, keepaspectratio]{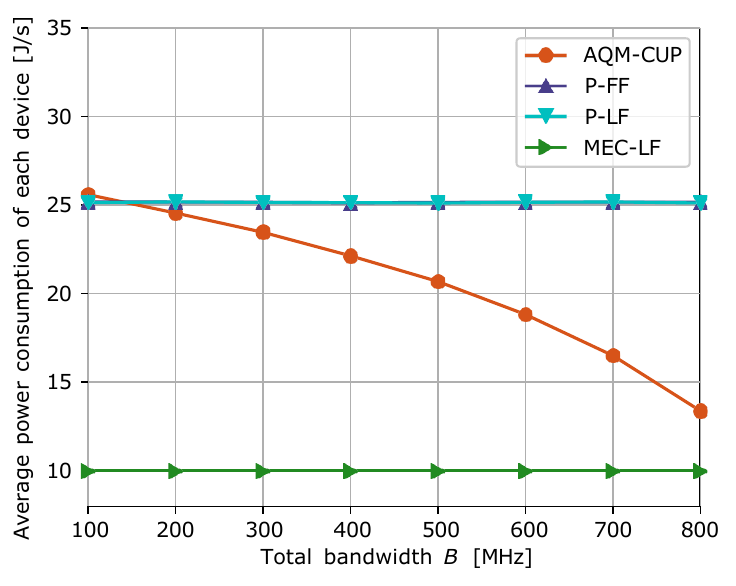}
\label{test_time_invariant_method_energy}
\end{minipage}
}
\caption{Performance of the AQM-CUP and time-invariant methods vs. total bandwidth with $F^{\nu}=70\,\mathrm{GHz}$, $\zeta=0.1$. }
\label{test_time_invariant_method}
\end{figure}

Fig. \ref{test_BF} illustrates the impact of communication and computational resources on the performance under the AQM-CUP algorithm. As total bandwidth and computing power increase, the metrics of sensor information age and device power consumption progressively improve. We can see that when the total bandwidth is 100 MHz, the average age of sensor information exceeds 40 ms (when $k^{\star} = k+L-1$, $\varkappa_{u,k}=40$ ms). As the total bandwidth increases, there is a notable initial decrease. This implies that increased bandwidth reduces dizziness caused by MTP timeout. A slower rate of reduction is observed in Fig. \ref{test_BF_age} as the average age of sensor information falls below 40 ms. At this point, nearly all frames meet the MTP threshold, and the decrease in sensor information is mainly attributable to  more recent prediction information. From Fig. \ref{test_BF_energy}, the average power consumption of each device decreases only slightly with the initial increase in bandwidth. This is because AQM-CUP prioritizes reducing MTP latency under limited bandwidth conditions, while reducing power consumption is comparatively less critical. As the total bandwidth continues to increase, a greater proportion of rendering tasks are executed at MEC, leading to a decrease in the power consumption of VR devices. 
    
Fig. \ref{test_algorithm} demonstrates the performance of the AQM-CUP. The advantages of the proposed algorithm in reducing the average age of sensor information become more pronounced when the total bandwidth is between $100$ MHz and $300$ MHz. This indicates that as bandwidth resource become more limited, a greater number of expired tiles are actively discarded from processing queues, making the discarded tiles information in AQM-CUP more influential. In Fig. \ref{test_algorithm_energy}, although AQM-CUP does not exhibit the same advantages in power consumption as it does in the age of sensor information, the average energy consumption remains relatively low across most bandwidths, consistent with the convergence results in Fig. \ref{training_performance_energy}.

To further illustrate the performance of the proposed AQM-CUP, three time-invariant methods are shown in Fig. \ref{test_time_invariant_method}. In these time-invariant methods, constant strategies are executed at each time slot: the total bandwidth $B$ and total GPU computation power $F^{\nu}$ are evenly distributed among users, while the rendering decisions differ as follows. (\romannumeral1) In P-FF, the foreground tile is locally rendered at the device, and the predicted background tile of the next time slot, $w^b_{u,k+1,k}$, is rendered in parallel at the MEC; (\romannumeral2) In P-LF, the parallel rendering method is the same as in P-FF, but the rendering index of the background tile is the last frame in the prediction window, $w^b_{u,k+L-1,k}$; (\romannumeral3) In MEC-LF, the rendering index of the background tile is $w^b_{u,k+L-1,k}$, and both the foreground and background tiles are rendered at MEC. 
In Fig. \ref{test_time_invariant_method_age}, P-LF outperforms other benchmarks in terms of the average age of sensor information. This is because rendering parallelization and rendering the last frame of the predicted background tile can mitigate MTP latency timeout by combining these two strategies. However, the performance of P-LF does not improve beyond $B = 300$ MHz (where the system meets the basic MTP threshold)  in Fig. \ref{test_time_invariant_method_age} and Fig. \ref{test_time_invariant_method_energy}, even with additional bandwidth increases. P-FF and MEC-LF can cause user dizziness, as shown in Fig. \ref{test_time_invariant_method_age}. MEC-LF has ultra-high bandwidth demand but offers the advantage of rendering all tiles at the MEC, resulting in the lowest device power consumption in Fig. \ref{test_time_invariant_method_energy}. Overall, AQM-CUP outperforms the time-invariant methods by minimizing the age of sensor information while reducing device power consumption as much as possible.


\begin{figure*}[t] 
\centering
\subfigure[Total rendering load in $Q^{r_e}_u$ and total data size in $Q^t_u$ are the same with the user index $u$.]{
\begin{minipage}[t]{0.3\linewidth}
\centering
\includegraphics[height=2.92cm, keepaspectratio]{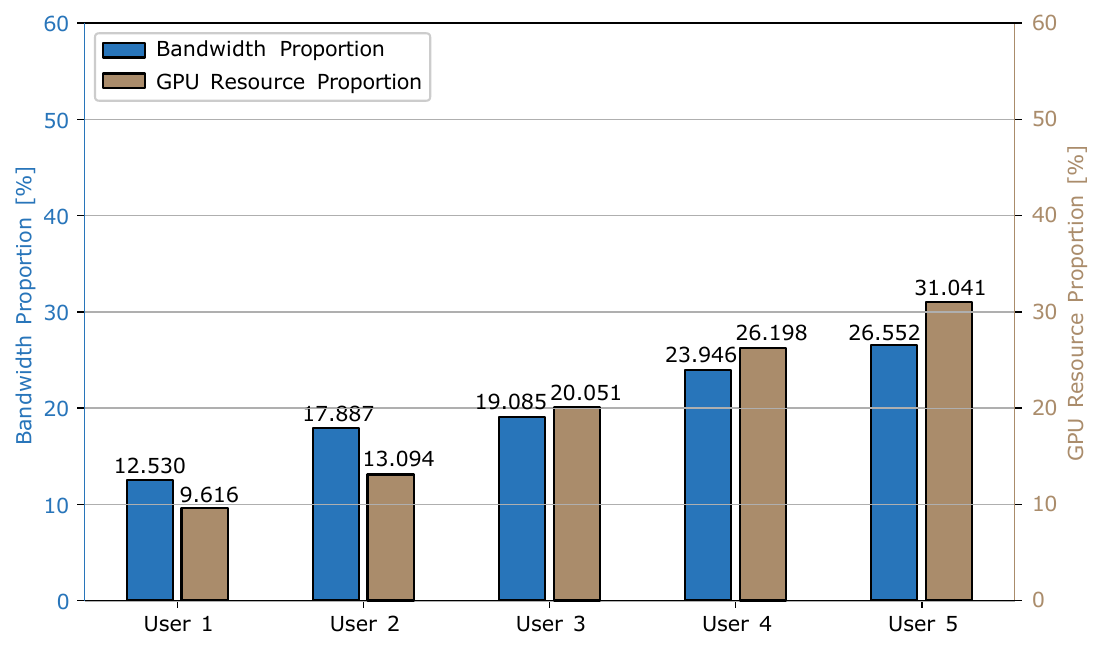}
\label{users_constant}
\end{minipage}
}
\hspace{0.15cm}
\subfigure[Total rendering load in $Q^{r_e}_u$ and total data size in $Q^t_u$ decrease with the user index $u$.]{
\begin{minipage}[t]{0.3\linewidth}
\centering
\includegraphics[height=2.92cm, keepaspectratio]{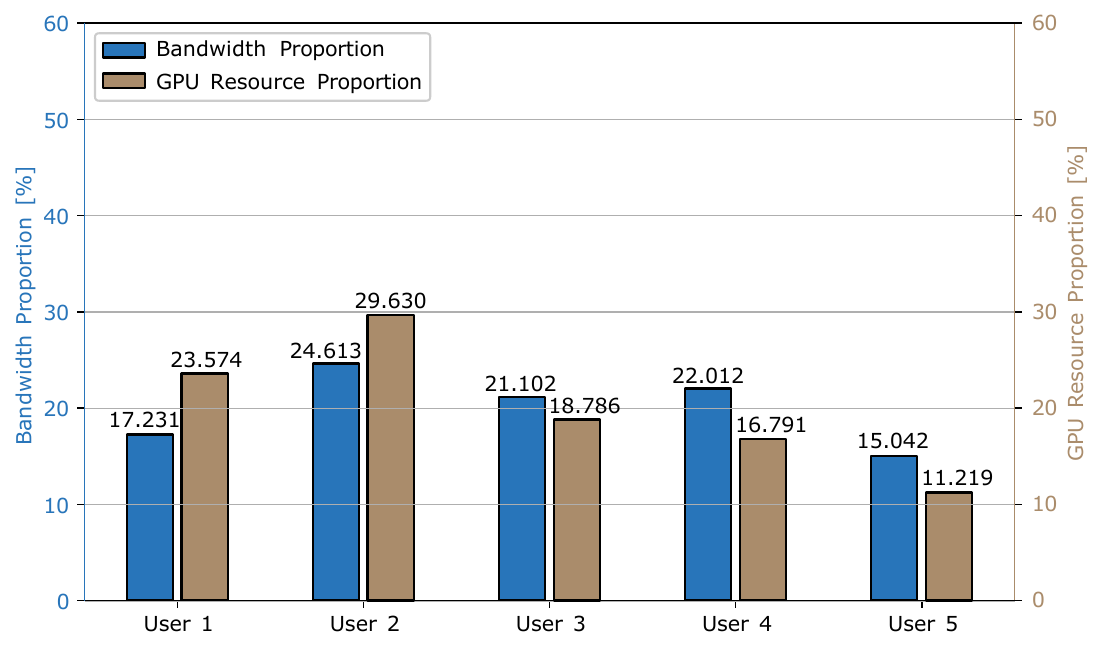}
\label{users_decrease}
\end{minipage}
}
\hspace{0.15cm}
\subfigure[Total rendering load in $Q^{r_e}_u$ and total data size in $Q^t_u$ increase with the user index $u$.]{
\begin{minipage}[t]{0.3\linewidth}
\centering
\includegraphics[height=2.92cm, keepaspectratio]{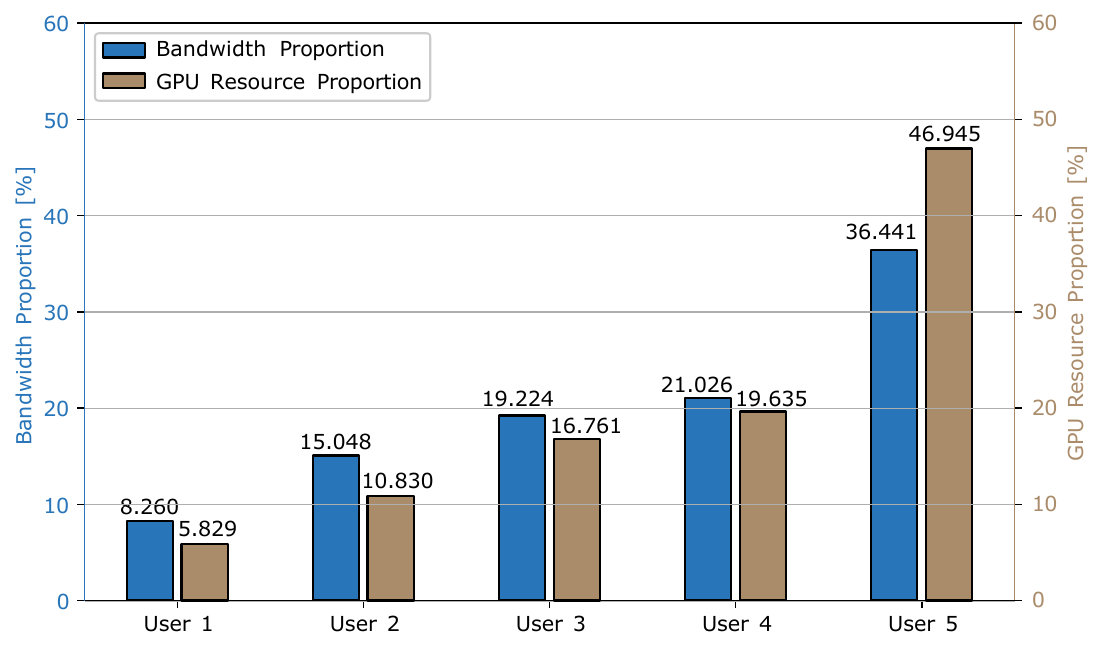}
\label{users_increase}
\end{minipage}
}
\hspace{1cm}
\caption{Resource allocation with $B=500$ MHz, $F^{\nu}$ = 70 GHz, $\zeta$ = 0.1, the channel state $h_{u,k}g_{u,k}$ improves as the user index $u$ increases. (a) The parallel rendering method is employed for all users, i.e., the rendering location of foreground tiles is on devices, while the rendering location of background tiles is on MEC; the rendering index of the background tile is $\{w^b_{u,k+3,k}\mid u\in \mathcal{U}\}$ for all users. (b) The parallel rendering method is executed for all users; the rendering indices of the background tile are $w^b_{1,k+4,k}$ and $\{w^b_{u,k+3,k}\mid u \in \{2, 3, 4, 5\}\}$. (c) The parallel rendering method is utilized for users $u\in\{1,2,3,4\}$, with both foreground and background tiles rendered at the MEC for user $5$; for all users, the rendering index of the background tile is $\{w^b_{u,k+3,k}\mid u\in \mathcal{U}\}$. }
\label{users}
\end{figure*}

Fig. \ref{users} verifies the effectiveness of the proposed algorithm in resource allocation. Fig. \ref{users_constant} shows that the user with a better channel state is allocated more resources by MEC to render and transmit data in the buffers. Comparing Fig. \ref{users_decrease} and Fig. \ref{users_constant}, the amount of data at MEC is more influential than channel conditions. This is evident as the user in Fig. \ref{users_decrease} with a larger rendering load and data size but worse channel condition is allocated more resources. Coordination between rendering decisions and resource allocation can be seen in Fig. \ref{users_decrease} and Fig. \ref{users_increase}. Although user 1 in Fig. \ref{users_decrease} has the largest amount of data in the MEC buffers to be processed, the allocated resources are not the highest, indicating that the larger rendering index of user 1 relaxes the MTP constraint. The effect of rendering locations is also notable. User 5 in Fig. \ref{users_increase} is entirely rendered by MEC, thereby obtaining more resources by significantly increasing the amount of data in the MEC buffer.

\section{Conclusion}\label{conclusion}
In this paper, we have design a wireless multi-user interactive VR with edge-device collaborative computing framework to address the ultra-low MTP threshold. We have formulated the optimization problem to minimize the age of sensor information and the power consumption of mobile devices while meeting the MTP constraint through rendering decisions and MEC resource allocation. We have proposed a safe RL algorithm, AQM-CUP to solve this optimization problem. Numerical results demonstrate that the proposed algorithm outperforms the considered baselines in terms of the training convergence and performance metrics.
\bibliographystyle{IEEEtran}
\bibliography{reference}

\begin{thebibliography}{10}
\providecommand{\url}[1]{#1}
\csname url@samestyle\endcsname
\providecommand{\newblock}{\relax}
\providecommand{\bibinfo}[2]{#2}
\providecommand{\BIBentrySTDinterwordspacing}{\spaceskip=0pt\relax}
\providecommand{\BIBentryALTinterwordstretchfactor}{4}
\providecommand{\BIBentryALTinterwordspacing}{\spaceskip=\fontdimen2\font plus
\BIBentryALTinterwordstretchfactor\fontdimen3\font minus
  \fontdimen4\font\relax}
\providecommand{\BIBforeignlanguage}[2]{{%
\expandafter\ifx\csname l@#1\endcsname\relax
\typeout{** WARNING: IEEEtran.bst: No hyphenation pattern has been}%
\typeout{** loaded for the language `#1'. Using the pattern for}%
\typeout{** the default language instead.}%
\else
\language=\csname l@#1\endcsname
\fi
#2}}
\providecommand{\BIBdecl}{\relax}
\BIBdecl

\bibitem{my_GC}
C.~Xu, Z.~Chen, M.~Tao, and W.~Zhang, ``Edge-device collaborative rendering for
  wireless multi-user interactive virtual reality in metaverse,'' in
  \emph{Proc. IEEE Global Commun. Conf. (GLOBECOM)}, Dec. 2023, pp. 3542--3547.

\bibitem{immersion_itu}
\textquotedblleft Draft New Recommendation ITU-R M.[IMT.FRAMEWORK FOR 2030 AND
  BEYOND] - Framework and overall objectives of the future development of IMT
  for 2030 and beyond", Document 5/131, June 2023.
  https://www.itu.int/md/R19-SG05-C-0131.

\bibitem{metaverse}
W.~Y.~B. Lim, Z.~Xiong, D.~Niyato, X.~Cao, C.~Miao, S.~Sun, and Q.~Yang,
  ``Realizing the metaverse with edge intelligence: A match made in heaven,''
  \emph{IEEE Wireless Commun.}, pp. 1--9, 2022.

\bibitem{mtp_latency}
S.~M. LaValle, A.~Yershova, M.~Katsev, and M.~Antonov, ``Head tracking for the
  oculus rift,'' in \emph{Proc. IEEE Int. Conf. Robot. Autom. (ICRA)}, 2014,
  pp. 187--194.

\bibitem{meta_quest3}
\textquotedblleft Meta Quest 3", https://www.meta.com/quest/quest-3/.

\bibitem{huawei_vr_white_paper}
Huawei iLab, \textquotedblleft Cloud VR Network Solution White Paper", 2018.
  https://www.huawei.com/minisite/pdf/ilab/cloud\_vr\_network\_solution\_\\white\_paper\_en.pdf.

\bibitem{apple_vision_pro}
\textquotedblleft Apple Vision Pro", https://www.apple.com/apple-vision-pro/.

\bibitem{Reliability_HetNets}
Z.~Gu, H.~Lu, P.~Hong, and Y.~Zhang, ``Reliability enhancement for vr delivery
  in mobile-edge empowered dual-connectivity sub-6 ghz and mmwave hetnets,''
  \emph{IEEE Trans. Wireless Commun.}, vol.~21, no.~4, pp. 2210--2226, 2022.

\bibitem{mmwave_compress}
S.~Gupta, J.~Chakareski, and P.~Popovski, ``mmwave networking and edge
  computing for scalable 360° video multi-user virtual reality,'' \emph{IEEE
  Trans. Image Process.}, vol.~32, pp. 377--391, 2023.

\bibitem{nature_sensors}
H.~C. Ates, P.~Q. Nguyen, L.~Gonzalez-Macia \emph{et~al.}, ``End-to-end design
  of wearable sensors,'' \emph{Nat. Rev. Mater.}, vol.~7, pp. 887--907, Nov.
  2022.

\bibitem{viewport_prediction}
Z.~Pan, Y.~Zhang, T.~Lin, and J.~Yan, ``Liveae: Attention-based and
  edge-assisted viewport prediction for live 360° video streaming,'' in
  \emph{Proc. Workshop Emerg. Multimed. Syst. (EMS)}, 2023, pp. 28--33.

\bibitem{body_Interaction_prediction}
X.~Peng, S.~Mao, and Z.~Wu, ``Trajectory-aware body interaction transformer for
  multi-person pose forecasting,'' in \emph{Proc. IEEE Conf. Comput. Vis.
  Pattern Recognit. (CVPR)}, 2023, pp. 17\,121--17\,130.

\bibitem{3C_sun}
Y.~Sun, Z.~Chen, M.~Tao, and H.~Liu, ``Communications, caching, and computing
  for mobile virtual reality: Modeling and tradeoff,'' \emph{IEEE Trans.
  Commun.}, vol.~67, no.~11, pp. 7573--7586, 2019.

\bibitem{infocom_placement}
L.~Wang, L.~Jiao, T.~He, J.~Li, and M.~Mühlhäuser, ``Service entity placement
  for social virtual reality applications in edge computing,'' in \emph{Proc.
  IEEE Int. Conf. Comput. Commun.}, 2018, pp. 468--476.

\bibitem{JSAC_dynamic_place}
Y.~Zhang, L.~Jiao, J.~Yan, and X.~Lin, ``Dynamic service placement for virtual
  reality group gaming on mobile edge cloudlets,'' \emph{IEEE J. Sel. Areas
  Commun.}, vol.~37, no.~8, pp. 1881--1897, 2019.

\bibitem{VRgame_Zhu}
Z.~Chen, H.~Zhu, L.~Song, D.~He, and B.~Xia, ``Wireless multiplayer interactive
  virtual reality game systems with edge computing: Modeling and
  optimization,'' \emph{IEEE Trans. Wireless Commun.}, vol.~21, no.~11, pp.
  9684--9699, 2022.

\bibitem{time_prediction}
X.~Wei, C.~Yang, and S.~Han, ``Prediction, communication, and computing
  duration optimization for vr video streaming,'' \emph{IEEE Trans. Commun.},
  vol.~69, no.~3, pp. 1947--1959, 2021.

\bibitem{FoV_Prediction_Cui}
L.~Zhao, Y.~Cui, Z.~Liu, Y.~Zhang, and S.~Yang, ``Adaptive streaming of 360
  videos with perfect, imperfect, and unknown fov viewing probabilities in
  wireless networks,'' \emph{IEEE Trans. Image Process.}, vol.~30, pp.
  7744--7759, 2021.

\bibitem{JSAC_prediction_rendering_transmission}
X.~Liu, Y.~Deng, C.~Han, and M.~D. Renzo, ``Learning-based prediction,
  rendering and transmission for interactive virtual reality in ris-assisted
  terahertz networks,'' \emph{IEEE J. Sel. Areas Commun.}, vol.~40, no.~2, pp.
  710--724, 2022.

\bibitem{Multicast_prediction}
C.~Perfecto, M.~S. Elbamby, J.~D. Ser, and M.~Bennis, ``Taming the latency in
  multi-user vr 360°: A qoe-aware deep learning-aided multicast framework,''
  \emph{IEEE Trans. Commun.}, vol.~68, no.~4, pp. 2491--2508, 2020.

\bibitem{cross_frame_prediction}
C.~Y. Chen and H.~Y. Hsieh, ``Cross-frame resource allocation with
  context-aware qoe estimation for 360° video streaming in wireless virtual
  reality,'' \emph{IEEE Trans. Wireless Commun.}, vol.~22, no.~11, pp.
  7887--7901, 2023.

\bibitem{JSAC_sample}
Z.~Meng, C.~She, G.~Zhao, and D.~De~Martini, ``Sampling, communication, and
  prediction co-design for synchronizing the real-world device and digital
  model in metaverse,'' \emph{IEEE J. Sel. Areas Commun.}, vol.~41, no.~1, pp.
  288--300, 2023.

\bibitem{task_oriented_meta}
Z.~Meng, K.~Chen, Y.~Diao, C.~She, G.~Zhao, M.~A. Imran, and B.~Vucetic,
  ``Task-oriented cross-system design for timely and accurate modeling in the
  metaverse,'' \emph{IEEE J. Sel. Areas Commun.}, vol.~42, no.~3, pp. 752--766,
  2024.

\bibitem{PAoI}
L.~Huang and E.~Modiano, ``Optimizing age-of-information in a multi-class
  queueing system,'' in \emph{Proc. IEEE Int. Symp. Inf. Theor. (ISIT)}, 2015,
  pp. 1681--1685.

\bibitem{back_fore}
Z.~Lai, Y.~C. Hu, Y.~Cui, L.~Sun, N.~Dai, and H.-S. Lee,
  ``\BIBforeignlanguage{English}{Furion: Engineering high-quality immersive
  virtual reality on today's mobile devices},''
  \emph{\BIBforeignlanguage{English}{IEEE Trans. Mobile Comput.}}, vol.~19,
  no.~7, pp. 1586--1602, 2020.

\bibitem{Q-vr}
C.~Xie, X.~Li, Y.~Hu, H.~Peng, M.~Taylor, and S.~L. Song, ``Q-vr: system-level
  design for future mobile collaborative virtual reality,'' in \emph{Proc. ACM
  Int. Conf. Archit. Support Program. Lang. Oper. Syst. (ASPLOS)}, 2021, pp.
  587--599.

\bibitem{segalopengl}
M.~Segal and K.~Akeley, ``The opengl{\textregistered} graphics system: A
  specification (version 4.6),'' 2022.

\bibitem{nature_dqn}
V.~Mnih, K.~Kavukcuoglu, D.~Silver \emph{et~al.}, ``Human-level control through
  deep reinforcement learning,'' \emph{Nature}, vol. 518, pp. 529--533, Feb.
  2015.

\bibitem{PPO}
J.~Schulman, F.~Wolski, P.~Dhariwal, A.~Radford, and O.~Klimov, ``Proximal
  policy optimization algorithms,'' \emph{arXiv:1707.06347}, 2017.

\bibitem{safeRL_openai}
A.~Ray \emph{et~al.}, ``{Benchmarking Safe Exploration in Deep Reinforcement
  Learning},'' \emph{OpenAI}, 2019,
  https://openai.com/research/benchmarking-safe-exploration-in-deep-reinforcement-learning.

\bibitem{CPO}
J.~Achiam, D.~Held, A.~Tamar, and P.~Abbeel, ``Constrained policy
  optimization,'' in \emph{Proc. Int. Conf. Mach. Learn.}, 2017, pp. 22--31.

\bibitem{CUP}
L.~Yang, J.~Ji, J.~Dai, L.~Zhang, B.~Zhou, P.~Li, Y.~Yang, and G.~Pan,
  ``Constrained update projection approach to safe policy optimization,'' in
  \emph{Proc. Adv. Neural Inf. Process. Syst. (NeurIPS)}, vol.~35.\hskip 1em
  plus 0.5em minus 0.4em\relax Curran Associates, Inc., 2022, pp. 9111--9124.

\bibitem{ATW}
J.~M.~P. van Waveren, ``The asynchronous time warp for virtual reality on
  consumer hardware,'' in \emph{Proc. ACM Symp. VR Softw. Technol.}\hskip 1em
  plus 0.5em minus 0.4em\relax New York, USA: Association for Computing
  Machinery, 2016, pp. 37--46.

\bibitem{3GPP_mmwave}
3GPP, \textquotedblleft ETSI TR 138 901 V16.1.0: 5G; Study on channel model for
  frequencies from 0.5 to 100 GHz (3GPP TR 38.901 version 16.1.0 release 16)",
  European Telecomm. Std. Institute (ETSI), 2020.
  https://www.etsi.org/deliver/etsi\_tr/138900\_138999/138901/16.01.00\_60/\\tr\_138901v160100p.pdf.

\bibitem{mmWave}
M.~Di~Renzo, ``Stochastic geometry modeling and analysis of multi-tier
  millimeter wave cellular networks,'' \emph{IEEE Trans. Wireless Commun.},
  vol.~14, no.~9, pp. 5038--5057, 2015.

\bibitem{FLOPs_Berkeley}
T.~A. Funkhouser and C.~H. Sequin, ``\BIBforeignlanguage{English}{Adaptive
  display algorithm for interactive frame rates during visualization of complex
  virtual environments},'' in \emph{\BIBforeignlanguage{English}{Proc. ACM
  Spec. Int. Group Comp. Graph. Interact. Tech. Conf. (SIGGRAPH)}}, Anaheim,
  CA, USA, 1993, pp. 247--254.

\bibitem{FLOPs}
M.~Wimmer and P.~Wonka, ``Rendering time estimation for real-time rendering,''
  in \emph{Proc. Eurographics Rendering Workshop (EGRW)}, Goslar, DEU, 2003,
  pp. 118--129.

\bibitem{CMDP}
E.~Altman, ``Constrained markov decision processes with total cost criteria:
  Lagrangian approach and dual linear program,'' \emph{Math. Method Oper.
  Res.}, vol.~48, no.~3, pp. 387--417, 1998.

\bibitem{GAE}
J.~Schulman, P.~Moritz, S.~Levine, M.~Jordan, and P.~Abbeel, ``High-dimensional
  continuous control using generalized advantage estimation,'' in \emph{Proc.
  Int. Conf. Learn. Represent. (ICLR)}, 2016.

\bibitem{Sketchfab}
\textquotedblleft Sketchfab", https://sketchfab.com/feed.

\end{thebibliography}

\end{document}